\documentclass[useAMS,usenatbib]{mnras}
\usepackage{epsfig, bm, times, aas_macros}
\usepackage[english]{babel}
\usepackage{amsmath}
\usepackage{amssymb}
\usepackage{graphicx}
\usepackage{color}

\newcommand{\seclab}{Section}
\newcommand{\tablab}{Table}

\newcommand{\figlab}{Figure}
\newcommand{\equlab}{Eq}
\newcommand{\secref}[1]{\seclab{} \ref{#1}}
\newcommand{\tabref}[1]{\tablab{} \ref{#1}}

\newcommand{\figref}[1]{\figlab{} \ref{#1}}
\newcommand{\figrefs}[1]{\figlab{s} \ref{#1}}
\newcommand{\eqr}[1]{\equlab{.} \eqref{#1}}
\newcommand{\eqrs}[1]{\equlab{s.} \eqref{#1}}

\newcommand{\code}[1]{\texttt{#1}}
\newcommand{\timt}{\times}
\newcommand{\bc}{art-2012-braith-cavecchi}
\newcommand{\cath}{art-2013-cavecchi-etal}
\newcommand{\cafi}{art-2015-cavecchi-etal}
\newcommand{\cael}{art-2011-cavecchi-etal}
\newcommand{\timm}{art-2000-tim}
\newcommand{\nuke}{\rm{n}}
\newcommand{\kaco}{\kappa_{\rm{c}}}
\newcommand{\tent}[1]{\times 10^{#1}}
\newcommand{\ro}{R_{\rm Ro}}
\newcommand{\citeple}[1]{\citeauthor{#1}, \citeyear{#1}}

\newcommand{\x}{x}
\newcommand{\y}{y}
\newcommand{\z}{z}

\newcommand{\uy}{{U_{\rm \y}}}
\newcommand{\uz}{{U_{\rm \z}}}

\newcommand{\Bf}{{B}}
\newcommand{\Bx}{{\Bf_{\rm \x}}}
\newcommand{\By}{{\Bf_{\rm \y}}}

\newcommand{\Bo}{ {\rm{\Bf}}_0}
\newcommand{\va}{v_{\rm A}}
\newcommand{\vf}{v_{\rm f}}

\newcommand{\vperp}{v_{\rm f\perp}}
\newcommand{\pt}{P_{\rm T}}
\newcommand{\pb}{P_{\rm B}}
\newcommand{\fa}{\phantom{0}}
\newcommand{\lr}{R_{\rm{R}}}

\newcommand{\tr}{\tau_{\rm R}}
\newcommand{\tb}{\tau_{\rm A}}
\newcommand{\tf}{\tau_{\rm fr}}
\newcommand{\tn}{\tau_{\rm n}}
\newcommand{\tbu}{\tau_{\rm burn}}
\newcommand{\impt}{\tr / \tf}
\newcommand{\crust}{\chi}
\newcommand{\scaH}{1.89}
\newcommand{\mx}{90}
\newcommand{\mz}{192}

\newcommand{\TF}{IGR J17480}
\newcommand{\TFf}{IGR J17480--2446}
\newcommand{\halftemp}{T_0 + \delta T /2}
\newcommand{\doubfig}[4]{%
  \includegraphics[width=0.45\textwidth]{#1}{#2}
  \hspace{\stretch{1}}
  \includegraphics[width=0.45\textwidth]{#3}{#4}\\
}
\newdimen \auxil
\newcommand{\brep}[2]{%
  \auxil = #1pt
  \ifdim\auxil=1pt
        {$\Bo = 10^{#2}$ G}
  \else
        {$\Bo = #1\tent{#2}$ G}
  \fi
}
\newcommand{\nrep}[1]{$\nu = #1$ Hz}
\newcommand{\bena}{B\'enard}
\newcommand{\alf}{Alfv\'en}
\newcommand{\slu}{art-2002-spit-levin-ush}
\newcounter{foota}
\newcommand{\tws}[1]{http://cococubed.asu.edu/code\_pages/#1}

\title[Magnetic flame fronts during Type I bursts]{Fast and slow
  magnetic deflagration fronts in Type I X-ray bursts}

\author[Y.\ Cavecchi]{Yuri Cavecchi$^{1}$\thanks{E-mail:
    y.cavecchi@uva.nl}, Yuri Levin$^{2,3}$, Anna L. Watts$^{1}$
  and Jonathan Braithwaite$^{4}$\\
  $^{1}$Astronomical Institute ``Anton Pannekoek, University of
  Amsterdam, Postbus 94249, NL-1090 GE Amsterdam, the
  Netherlands\\
  $^{2}$ Monash Center for Astrophysics and School of Physics,
  Monash University, Clayton, VIC 3800, Australia\\
  $^{3}$ Physics Department and Columbia Astrophysics Laboratory,
  Columbia University, 538 West 120th Street, New York, NY 100
  27, USA\\
  $^{4}$ Argelander Institut f\"ur Astronomie, Universit\:at Bonn,
  Auf dem H\"ugel 71, D-53121 Bonn, Germany}

\pagerange{\pageref{firstpage}--\pageref{lastpage}} \pubyear{}

\begin{document}
\maketitle
\label{firstpage}

\begin{abstract}
Type I X-ray bursts are produced by thermonuclear runaways that
develop on accreting neutron stars. Once one location ignites, the
flame propagates across the surface of the star. Flame propagation is
fundamental in order to understand burst properties like rise time and
burst oscillations. Previous work quantified the effects of rotation
on the front, showing that the flame propagates as a deflagration and
that the front strongly resembles a hurricane. However the effect of
magnetic fields was not investigated, despite the fact that magnetic
fields strong enough to have an effect on the propagating flame are
expected to be present on many bursters. In this paper we show how the
coupling between fluid layers introduced by an initially vertical
magnetic field plays a decisive role in determining the character of
the fronts that are responsible for the Type I bursts. In particular,
on a star spinning at 450 Hz (typical among the bursters) we test seed
magnetic fields of $10^{7} - 10^{10}$ G and find that for the medium
fields the magnetic stresses that develop during the burst can speed
up the velocity of the burning front, bringing the simulated burst
rise time close to the observed values. By contrast, in a magnetic
slow rotator like \TFf, spinning at $11$ Hz, a seed field $\gtrsim
10^9$ G is required to allow localized ignition and the magnetic field
plays an integral role in generating the burst oscillations observed
during the bursts.
\end{abstract}


\section{Introduction}
\label{sec:intro}
The Type I Bursts are X-ray flashes that last for tens to hundreds of
seconds and are extremely bright, being able to produce up to
$10^{38}$ erg/s. They are observed in approximately 100 sources, which
are all low mass X-ray binaries where the accreting compact object is
a neutron star (NS). The bursts are thermonuclear runaways that ignite
in the accreted layers on the surface of the neutron stars
\citep[see][for
  reviews]{rev-2003-2006-stro-bild-book,rev-2008-gal-mun-hart-psal-chak}. It
is usually thought that ignition takes place at one position and then
propagates across the surface \citep{art-1982-shara}. Thus, the
observed lightcurve is the result of the convolution of the local
emission evolution at every point on the surface and the propagation
of the flame.

Type I burst provide a very promising means to extract information
about the underlying NS, such as mass and radius that could put
constraints on the equation of state of the matter in the core of the
star \citep{rev-2003-2006-stro-bild-book,rev-2013-miller-arxivpaper}.
However, many questions still remain open about the bursts, for
example regarding what the most likely ignition site is, what nuclear
reactions take place and how does the flame spreads over the
surface. In order to disentangle the information we seek from the
lightcurve, it is fundamental to understand how the flame propagates.
The propagation mechanism is the focus of this paper.

Early discussion about possible mechanisms and the resulting flame
velocity can be found in \citet{art-1981-wall-woss},
\citet{art-1982-fryx-woos-a}, \citet{art-1982-fryx-woos-b},
\citet{art-1995-bild} and \citet{art-2001-zing-etal} (see also
\citealt{art-2012-sim-gryaz-etal-a,art-2012-sim-gryaz-etal-b}).
\cite{\slu}\defcitealias{\slu}{SLU02} (from now on SLU02) suggested
that localized ignition relies on confinement of the hot fluid. If not
confined, the hot fluid would spread out over the surface, dissipate
its heat and as a consequence no ignition would take place. In the
presence of confinement, on the other hand, the heat would not
dissipate and the flame would spread thanks to some heat transport
mechanism at work between the hot and the cold fluid. Since most of
the bursters spin faster than $\sim 100$ Hz, the authors suggested
that the Coriolis force could provide the necessary confinement
creating hurricane-like systems. The horizontal scale of both the
ignition and the propagating front is that of a Rossby deformation
radius, $\lr = \sqrt{gH}/2\Omega$, where $g$ is the gravitational
acceleration, $H$ the scale height of the fluid and $\Omega$ the
angular velocity of the star. This was convincingly demonstrated by ab
initio numerical simulations of \citet{\cath,\cafi}. \citetalias{\slu}
also discussed the frictional coupling between the top and bottom
layers and this is very important for this work (see below), but
limited themselves to parametrizing its effect on the flame
velocity. The simulations of \citet{\cath,\cafi} obtained propagation
time scales in good agreement but generally somewhat longer than the
observed rise times. Furthermore, they found that the principal
mechanism responsible for the heat transport was the top-to-bottom
conduction across the flame front. These simulations were performed
without any magnetic field, which we show in this paper dramatically
changes the velocity and the character of the flame.

Moreover, \citet{\cael} discussed a very peculiar source, \TFf{} (from
now on \TF{}, see also \citeple{art-2011-manu-mhric},
\citeple{art-2011-motta}), which is spinning at approximately 11
Hz. This source is very important, since at 11 Hz the Coriolis force
is not effective in confining the hot fluid, the Rossby radius being
greater than the star's radius. Nonetheless, \TF{} shows bursts and
burst oscillations. The latter are fluctuations in the burst
luminosity that are caused by an asymmetry in the emitting pattern
modulated by the rotation of the star (first discovered by
\citealt{art-1996-stro-etal}, they are seen in many bursters at
different rotation rates, see \citealt{rev-2012-watts} for a
review). \TF{} is an accretion powered millisecond pulsar:
fluctuations are also seen in the persistent emission, which implies
that the accretion flow is channelled by a significant magnetic field
\citep[see][for a review on the behaviour of accretion powered X-ray
  pulsars]{rev-arx-2012-pat-watts}. Finally, the burst oscillations
and accretion powered oscillations of \TF{} are in phase. Based on
this evidence, \citet{\cael} proposed that the magnetic field could be
the source of flame confinement in this case and estimated that a
field $\gtrsim 10^{9}$ G was necessary.

In this paper we explore the effects of the magnetic field on the
flame propagation of Type I Bursts in order to establish its
interaction with the Coriolis force. We explore conditions that apply
to the bursting X-ray pulsars, where the extra coupling between the
layers provided by the magnetic field threading them is likely to
affect the speed of flame propagation, increasing it and thus bringing
our results to better agreement with the observations. We also
dedicate a section specifically to the case of slow rotators like
\TF{} where the magnetic field is needed to provide the extra
confinement that the Coriolis force is not able to provide. In
\secref{sec:sims} we present the results of analytical calculation and
the numerical simulations and in \secref{sec:discussion} we draw our
conclusions.

\section{The interaction between the magnetic field and the burning front}
\label{sec:sims}
Here we will describe the results of our analytical calculations and
numerical simulations, but first we summarize the numerical setup
common to all runs.

\subsection{Numerical setup}
\label{sec:numerical}
We use the code described in \citet{\bc} and \citet{\cath}. An
important assumption behind our simulations is hydrostatic
equilibrium. This approach is particularly well suited for fluids
whose horizontal stretch is much greater than the vertical extent, as
in the case of the accreted layers of NSs. We use the so called
$\sigma$ coordinate system \citep[see][]{art-1974-kasa}, a kind of
pressure coordinate system. In the $\sigma$ coordinate system, the
\emph{physical height} of the fluid is fixed at the bottom to a
constant value ($0$ cm in our case), while for every other grid point
the physical height will change according to the fluid evolution. On
the other hand, the \emph{pressure} at the top is held constant
throughout the simulation, while the bottom pressure will change
according to the motion of the fluid. Of the various coordinate
systems used in atmospheric physics, this is the logical choice in
this context, as the computational domain can expand and contract to
arbitrary degree in the vertical direction in response to the heating
and cooling of the ocean.\footnote{Due to the change of coordinates,
  the equations for the magnetic part described in \citet{\bc} should
  contain terms proportional to powers of the horizontal derivatives
  of the height (the terms $\partial_h \phi$ in \citealt{\bc}). We
  neglect these since they turn out to be small due to the extreme
  aspect ratio of our simulations.}

The code uses finite differences and an explicit third-order
low-storage Runge-Kutta scheme \citep{art-1980-will} as the time
stepper. The time step is limited by a Courant condition
\citep[see][]{\bc}. In our simulations the most limiting factors to
the timestep are fluid motion and the magnetic field
evolution. Spatial derivatives and integrals are fifth order accurate,
while interpolations are sixth order (for details see
\citealt{art-1992-lele} and \citealt{\bc}). Variables are evaluated on
a staggered grid: thermodynamic variables are defined at the centers
of each grid cell, while velocities and magnetic field components are
face centred. The staggered grid ensures good conservation of mass,
energy, momentum and divergence of magnetic field \citep[][for the
  simulations in this paper, the average $\nabla B / \Bo$ is at
  maximum $10^{-12}$]{\bc}. The relation between density, pressure,
temperature and composition is calculated with the routine
\code{helmeos}\footnote{Publicly available at \tws{eos.shtml}.} of
\citet{art-2000-tim-swes} where electrons can be arbitrarily
degenerate and relativistic, ions are considered ideal gas and
radiation pressure is also taken into account.

In this paper we use Cartesian coordinates, as in \citet{\cath}, but
we set our coordinates system such that $\x$ is pointing in the
vertical direction, $\z$ in the horizontal direction parallel to plane
of the simulations (the direction of flame propagation) and $\y$ is
pointing out of the plane. We also neglect one horizontal direction,
assuming symmetry along the $\y$ direction. This configuration will
help us elucidate most clearly the role of magnetic field in setting
the fronts propagation speed. Vertically, our fluid is
\emph{initially} between $\pt=10^{22}$ erg cm$^{-3}$ and
$\pb=e^{\scaH}\tent{22}$ erg cm$^{-3}$, where $\pt$ and $\pb$ are the
pressure at the top and at the bottom of the simulation. The value of
\scaH{} in the exponent for $\pb$ implies that we are simulating
approximately \scaH{} scale heights.  Horizontally, the domain is
$6\tent{5}$ cm wide.

As in previous works, we set up an initial temperature profile which
is meant to trigger the burst. It is independent of the vertical
coordinate, but it has a horizontal dependence which is:
\begin{equation}
\label{eq:Tpert}
T = T_0 + \frac{\delta T }{1 + \exp[(\z - \z_0)/\delta\z]}
\end{equation}
The parameters we use in this paper are $T_0=2\tent{8}$ K, $\delta
T=2.81\tent{8}$ K, $\z_0 = 9\tent{4}$ cm and $\delta\z=3.6\tent{4}$
cm. $\z_0$ corresponds to the position where the temperature is
$\halftemp$, while $\delta\z$ is approximately the distance over which
the temperature transitions from its maximum $T_0 + \delta T$ to its
minimum $T_0$.

We only consider helium burning via the triple-$\alpha$ reaction
\citep{art-2000-cum-bild}\footnote{We neglect electron screening. For
  comparison, a run which uses the full reaction rate including the
  screening factor \citep[as in][]{art-1987-fushi-lamb-a} produces a
  flame velocity which is only 1.14 times faster. Since the focus of
  this paper is on the effects of the magnetic field we keep the
  simplified form \eqr{eq:reaction}.}
\begin{equation}
\label{eq:reaction}    
Q_{\nuke}=5.3\timt10^{18}\rho_5^2 \left(\frac{Y}{T_9}\right)^3 
e^{-4.4/T_9}\; \textrm{erg g$^{-1}$ s$^{-1}$},
\end{equation}
and we simulate a fluid which is a mixture of helium and carbon.

We assume that the bottom boundary is not moving (no-slip
condition). This is based on the consideration that the density of the
burning layer becomes much smaller than the density of the ashes on
which the layer rests. Therefore we allow for a buffer zone of pure
carbon at the bottom of the simulation which is not affected by the
flame propagation. In order to do so the initial composition is
constant horizontally, but the helium fraction decreases downwards
according to
\begin{equation}
Y = 1 - \frac{1}{1 + \exp[-(\sigma - 0.8)/1.6\tent{-2}]}
\end{equation}
($\sigma=0$ at the top and $\sigma=1$ at the bottom).  Erring on the
side of caution, we also impose a damping force to prevent motion in
the carbon layer near the boundary. Reassuringly, in later tests it
turned out that switching off the damping force makes very little
difference\footnote{We ran some simulations without the damping force
  and obtained qualitatively similar results, thus we conclude that
  this force does not affect the main conclusions of the paper.}. We
calculate the acceleration terms $\partial_t\uz$ and $\partial_t\uy$
and then multiply them by
\begin{equation}
\crust = \left \{\begin{array}{ccl} 
1 & \sigma < 0.82\\ 
\cos^4 \left(\frac{\sigma - 0.82}{0.1}\frac{\pi}{2}\right) 
& 0.82 < \sigma < 0.92 \\ 
0 & \sigma > 0.92
\end{array}\right.
\end{equation}
The choice of this functional form for $\crust$ is made to ensure
smoothness and the fact that the damping only acts within the buffer
zone of carbon.

In our initial configuration the magnetic field is purely vertical and
has a homogeneous value $\Bo$, which is different for different
simulations. The gravitational acceleration is $g=2\tent{14}$ cm
s$^{-2}$ and we use the plane-parallel approximation and a constant
Coriolis parameter $f=2\Omega$ (the $f$-plane approximation). The
fluid is initially at rest, with the horizontal velocities $\uz=\uy=0$
cm s$^{-1}$. The physical conduction is calculated with an opacity
fixed to $\kaco = 0.07$ cm$^2$ g$^{-1}$ \citep[see details
  in][]{\cath}.

On the horizontal components of the velocity we impose symmetric
boundary conditions at the vertical boundaries and reflective
conditions at the horizontal boundaries\footnote{Symmetric conditions
  imply that the values of the physical quantity are the same on
  either side of the boundary, while antisymmetric conditions imply
  that the values change sign. Reflective conditions imply that the
  velocities have antisymmetric conditions while thermodynamical
  quantities like temperature and density are symmetric. In this way
  the boundaries on the left and right act like rigid walls that
  confine the fluid inside the simulation domain.}. For the horizontal
components of the magnetic field we use antisymmetric boundary
conditions both horizontally and vertically, while for the vertical
component we use symmetric boundary conditions. By this choice, the
field is forced to be perpendicular to the bottom and top surfaces. In
all simulations presented here, we use horizontal and vertical
resolutions of \mz{} and \mx{} grid points respectively. Our
horizontal spatial resolution is therefore $3215$ cm, as in
\citet{\cath}, which is greater than the scale height. This
illustrates the suitability of the hydrostatic scheme. Our numerical
hyperdiffusive scheme has viscosity parameters $\nu_1=0.03$ and
$\nu_2=0.5$ \citep[see][for details]{\bc}. The corresponding values
for the temperature and the helium fraction $Y$ numerical
diffusivities are taken to be $1$ \% of these\footnote{However, Runs
  2, 3 and 7 needed a slight increase in the numerical thermal
  diffusivity to ensure numerical stability after some time during the
  simulation, while Run 5 needed it since the beginning due to the
  fast fluid motions (see \tabref{tab:runs} for the parameters of
  each run).}. We also use an explicit hyperdiffusive scheme for the
magnetic field to ensure stability and avoid zig-zags, it is
calculated as a multiple of the one used for the velocity \citep{\bc}.

\subsection{Analytical calculations}
\label{sec:results-calc}

\begin{table*}
\centering
\begin{tabular}{|r|c|c|c|c|c|c|c|c|}
Run&$\Bo\;(\textrm{G})$&$\nu\;(\textrm{Hz})$&$\tr$ (s)&$\tf$ (s)&$\lr$ (cm)&$\vf$ (cm/s)& Strength\\ 
\hline 
0 & ---              & 450  & $1.77\tent{-4}$&---            &$3.54\tent{4}$&$2.06\tent{5}$ &  ---        \\
\hline
1 & $1\tent{7\fa}$   & 450  & $1.77\tent{-4}$&$1.26\tent{-1}$&$7.91\tent{4}$&$7.28\tent{5}$ & Weak        \\
2 & $1\tent{8\fa}$   & 450  & $1.77\tent{-4}$&$1.26\tent{-3}$&$7.91\tent{4}$&$1.61\tent{6}$ & Intermediate\\
3 & $1\tent{9\fa}$   & 450  & $1.77\tent{-4}$&$1.26\tent{-5}$&$7.91\tent{4}$&$4.78\tent{5}$ & Intermediate-Strong\\
4 & $1\tent{10}  $   & 450  & $1.77\tent{-4}$&$1.26\tent{-7}$&$7.91\tent{4}$&$1.29\tent{5}$ & Strong      \\
\hline
5 & $1\tent{7\fa}$   & \fa11& $7.23\tent{-3}$&$1.26\tent{-1}$&$3.24\tent{6}$&---            & Weak        \\
6 & $1\tent{8\fa}$   & \fa11& $7.23\tent{-3}$&$1.26\tent{-3}$&$3.24\tent{6}$&---            & Intermediate - Strong\\
7 & $1\tent{9\fa}$   & \fa11& $7.23\tent{-3}$&$1.26\tent{-5}$&$3.24\tent{6}$&$4.59\tent{5}$ & Strong      \\
8 & $1\tent{10}  $   & \fa11& $7.23\tent{-3}$&$1.26\tent{-7}$&$3.24\tent{6}$&$1.33\tent{5}$ & Strong      \\
\hline
\end{tabular}
\caption{Values of the initial purely vertical magnetic field $\Bo$,
  the spin frequency $\nu = \Omega/2\pi$, the rotational $\tr=1/f$ and
  effective frictional (magnetic) timescales \eqr{eq:tfric} defined in
  \secref{sec:results-calc}, the Rossby radius $\lr=\sqrt{gH}\tr$ and
  the flame spread velocity $\vf$ for the main runs discussed in the
  text. The last column reports the rough classification used in
  \secref{sec:results-num}. The estimates for $\tf$ are obtained
  substituting the value of $\Bo$ in \eqr{eq:tbval}, using $\tbu=0.1$
  s and setting the numerical constants to $1$. For $\lr$ we use
  $g=2\tent{14}$ cm s$^{-2}$ and $H=10^{3}$ cm. $\vf$ is the value
  given by a \emph{linear} fit to the horizontal position of maximum
  burning as a function of time.}
\label{tab:runs}
\end{table*}

As shown in \citep{\cath}, when the flame propagates in the absence of
a magnetic field, the burning front is located along an inclined but
nearly horizontal surface that separates as yet unburnt fuel from
actively burning material (see e.g. \figref{fig:bf0wa}, left panel). A
strong vertical shear is set up within the front. In an atmosphere
without friction the horizontal velocity is in (quasi) geostrophic
balance with the horizontal pressure gradient. As a consequence, the
vertical shear of the velocity is given by the thermal wind relation
(a hurricane-like structure, see \citealt{book-1987-Pedlo} and
\citetalias{\slu} and \figref{fig:bf0wa} right panel):
\begin{equation}
  {\partial v_y\over \partial \x}={g\over f}
  {\partial\log\rho\over \partial \z}
\end{equation}   
As the horizontal gradient of the density is concentrated within the
burning front, so is the vertical shear of the horizontal fluid
velocity. When a magnetic field is present, this shear directly
couples to the pre-existing vertical magnetic field $B_\x=\Bo$ and
generates the horizontal field $B_\y$ via ${\rm{d}}B_\y/{\rm{d}}t=B_\x
\partial v_\y/\partial \x$; here $d/dt$ is the Lagrangian derivative
for the fluid element coming through the front.
 
The horizontal field generates magnetic stress that back-reacts on the
shear flow. The result is that the geostrophic balance is (partially)
broken, which can lead to faster penetration of the hot burning fluid
in the top layers of the unburnt fuel. The acceleration of the front
propagation due to the mechanical top-bottom momentum exchange was
noted in \citetalias{\slu} who introduced a phenomenological timescale
$\tf$ for the frictional coupling and showed that the front velocity
reaches maximum when $\tau_{\rm fr}\sim 1/f$. Here we study the effect
of the back-reaction of the flow-generated magnetic field and show
that it does indeed accelerate the front propagation for a certain
range of magnetic field strengths (compare to sections 2.3 and 3.7 of
\citetalias{\slu}).

Let $\vec{v}_{\rm h}$ be the horizontal velocity at the top side of
the burning front. The vertical shear is, to order of magnitude,
$v_h/h$ where $h$ is the vertical width of the front. The horizontal
magnetic field generated in a fluid element upon its passage through
the front is
\begin{equation}
\vec{B}_{\rm h}\sim B_\x {\tn \vec{v}_{\rm h}\over h}.
\end{equation}
Here $\tn$ is the nuclear burning time which is also the
characteristic time for a fluid element to cross the front. The
horizontal magnetic field causes horizontal acceleration of the top
velocity of the burnt-fuel
\begin{equation}
\vec{a}_{\rm mag}\sim-{B_\x \vec{B}_{\rm h}\over 4\pi \rho H},
\end{equation}
where $\rho$ and $H$ are the density and scale height of the
\emph{burning fluid} \citep{art-2009-heng-spit}. The overall
horizontal acceleration of the burning fluid is caused by three
physically distinct force fields: the horizontal pressure gradient,
the Coriolis force, and the magnetic back-reaction force:
\begin{equation}
  \vec{a}_h=\alpha_1 g{H\over L}\vec{e}_\z-\vec{f}\times\vec{v}_h-
  \alpha_2{\Bo^2\over 4\pi\rho H}{\tn\over h}\vec{v}_h,
\label{forcebalance}
\end{equation}
where $\alpha_1$ and $\alpha_2$ are numerical coefficients of order
$1$, $L$ is the horizontal extent of the front surface and we used
$\Bx=\Bo$, the seed field. To complete this equation, we note that
first,
\begin{equation}
\vec{a}_h\sim \alpha_0\vec{v}_h/\tbu,
\label{acceleration}
\end{equation}
and secondly,
\begin{equation}
L\sim \tbu\vec{v}_h\cdot \vec{e}_\z.
\label{length}
\end{equation}
Here $\tbu\sim \tn H/h$ is the timescale for the front to burn
vertically through a patch of the ocean and $\alpha_0$ is a constant
of order $1$. \eqr{acceleration} reflects the fact that the fluid has
time $\tbu/\alpha_0$ to accelerate from zero to $\vec{v}_h$ as the
front burns vertically through the ocean, while \eqr{length} reflects
the fact that it is the velocity component $\vec{v}_h\cdot \vec{e}_\z$
that is responsible for the hot fluid covering the unburnt cold fluid.

\eqrs{forcebalance}, \eqref{acceleration} and \eqref{length} form a
closed system that can be solved for the two horizontal components of
velocity $\vec{v}_h\cdot \vec{e}_\z$ and
$\vec{v}_h\cdot\vec{e}_\y$. The $\z$-component is representative of
the horizontal velocity of the burning front
\citepalias[compare][equation 34]{\slu}; it is given by
\begin{equation}
  \vf\sim\vec{v}_h\cdot\vec{e}_\z=
  \left[{{\alpha_1 gH \over \tbu}\frac{1/ \tau}{f^2 + 1/\tau^2}}\right]^{1/2}.
\label{vfront}
\end{equation}
Here 
\begin{equation}
\tau=\left(\alpha_2{\tbu\over \tb^2}+\frac{\alpha_0}{\tbu}\right)^{-1},
\label{tau}
\end{equation}
where
\begin{equation}
\tb = H/\va = {\sqrt{4\pi\rho} H/\Bo}
\end{equation}
is the \alf{} crossing timescale and $\va = \Bo /2 \sqrt{\pi\rho}$ the
\alf{} speed. It can be seen that the effective frictional time is
given in our case by
\begin{equation}
  \label{eq:tfric}
  \tf = \frac{\tb^2}{\alpha_2\tbu}
\end{equation}

Let us consider several limits of this equation. In the non-magnetic
limit, so that $\tb\gg \tbu$, the front velocity becomes (using $1/f
\ll \tbu$)\footnote{Later, we will consider a case when \nrep{11}. In
  this case $1/f\sim7\tent{-3}$ s, which still satisfies this
  condition.}

\begin{equation}
  \vf\sim \sqrt{\alpha_1\alpha_0}{\sqrt{gH}\over f\tbu},
  \label{vlow}
\end{equation}
in agreement with \citet{\cath} and \citetalias{\slu}. Note that $\vf$
is independent of the magnetic field coupling. However, for higher
magnetic fields
\begin{equation}
  \tb=0.01\left({H\over 10^3\hbox{ cm}}\right)
  \left({\rho\over 10^5 \hbox{ g}/\hbox{cm}^3}\right)^{1/2}
  \left({\Bo\over 10^8\hbox{ G}}\right)^{-1} \hbox{s}
  \label{eq:tbval}
\end{equation}
may well be much smaller than $\tbu\sim \hbox{few}\times 0.1\hbox{s}$
so that $\tau\sim\tf$. So the next relevant comparison is that of the
timescale $\tf$ with $\tr=1/f$. In the strong-field case where $\tf\ll
\tr$, which corresponds to
\begin{eqnarray}
  \Bo\gg \bar{\Bo}&=&2.5\times 10^8
  \left({H\over 10^3\hbox{ cm}}\right)
  \left({\rho\over 10^5\hbox{ g}/\hbox{cm}^3}\right)^{1/2}\times\nonumber\\
  & & \left({\nu\over 400\hbox{ Hz}}\right)^{1/2}
      \left({\alpha_2\tbu\over 0.1\hbox{ s}}\right)^{-1/2}\hbox{G},
\label{BZcr}
\end{eqnarray}
the speed of the front becomes independent of the rotation rate:
\begin{equation}
  \vf\sim \sqrt{gH}\sqrt{\frac{\alpha_1\tf}{\tbu}}
  \sim \sqrt{gH}\frac{\tb}{\tbu}\sqrt{\frac{\alpha_1}{\alpha_0}}.
  \label{vhigh}
\end{equation}
Finally, in the intermediate regime of $\Bo\sim \bar{\Bo}$, i.e.
$\tf\sim\tr(\ll\tbu)$ the magnetic field strongly accelerates the
speed of the flame propagation, reaching the maximum at
$\Bo=\bar{\Bo}$:
\begin{equation}
  \hbox{max}\left(\vf\right)=\sqrt{\alpha_1/2}{\sqrt{gH}\over \sqrt{f \tbu}}.
  \label{vmax}
\end{equation}
The analytic arguments of this section are confirmed by the direct
numerical simulations described below.

\subsection{Numerical simulations}
\label{sec:results-num}

We simulated different configurations varying the initial vertical
magnetic field intensity and the spin frequency of the star. We
consider $\Bo$ in the range $10^7 - 10^{10}$ G (broadly covering the
range of magnetic fields in accretion powered X-ray pulsars
\citealt{art-2015-muk-bult-klis-bhatta}) and \nrep{450} and \nrep{11};
a summary of the parameters for each run is reported in
\tabref{tab:runs}. We divide our simulations into three rough
categories, which we call \emph{strong}, \emph{intermediate} and
\emph{weak} magnetic field based on the $\impt$ ratio and the
development of the hurricane structure. We discuss these cases
separately, comparing them to the case of Run 0, which has no magnetic
field present and is rotating at \nrep{450}. This simulation evolves
as those discussed in \citet{\cath}. In particular, it has a well
defined hurricane structure\footnote{More precisely, the result of our
  simulations is an infinitely long roll, due to the use of flat
  Cartesian coordinates, which is the analogue to a hurricane
  structure.} identifiable by the thermal wind (see
\figref{fig:bf0wa}, right panel). It also has a well defined burning
region without turbulence (\figref{fig:bf0wa}, left panel).

\subsubsection{Strong magnetic field ($10^{10}$ {\rm{G}})}
\label{sec:strong}

\begin{figure*}
  \centering
  \doubfig{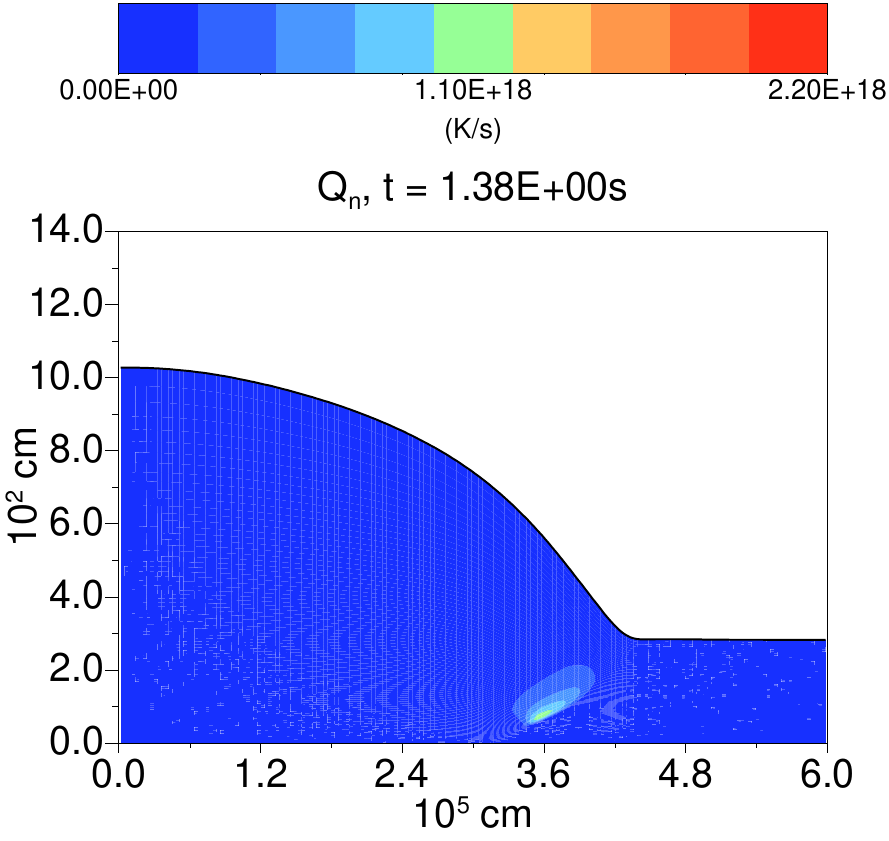}{}{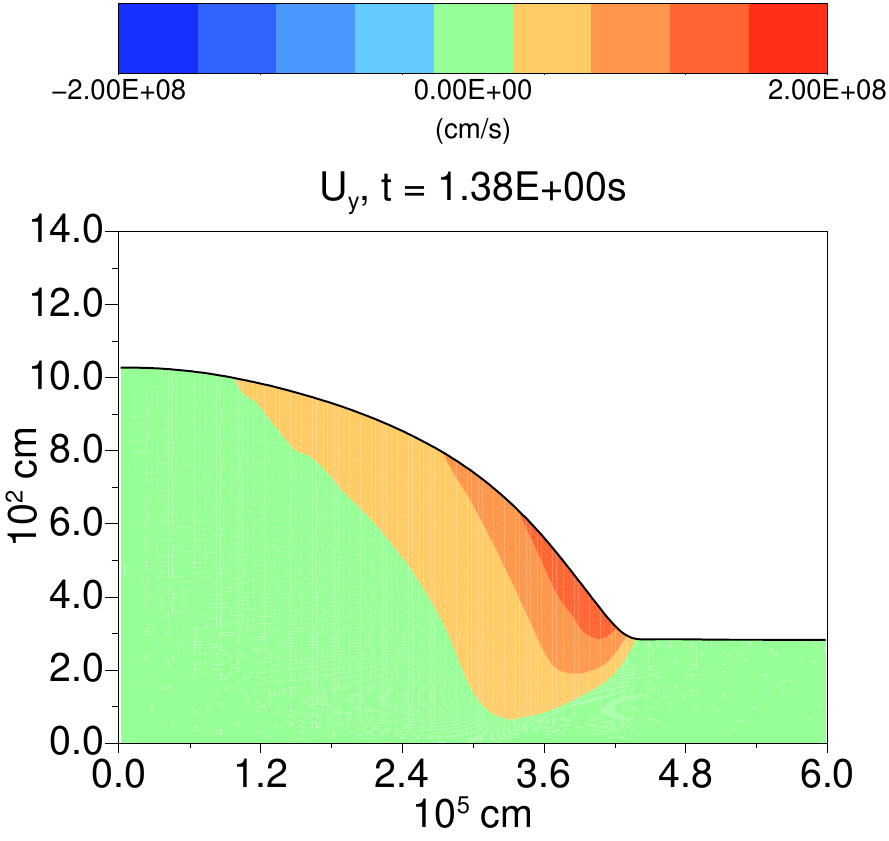}{}
  \caption{Burning rate (\textbf{left}) and velocity in the \y{}
    direction (perpendicular to the plane of the simulation -
    \textbf{right}) for the the non magnetic case, Run 0, with
    \nrep{450}. The flame along the hot-cold fluid interface without
    turbulence and the thermal wind are clearly visible.}
  \label{fig:bf0wa}
\end{figure*}

\begin{figure*}
  \centering
  \doubfig{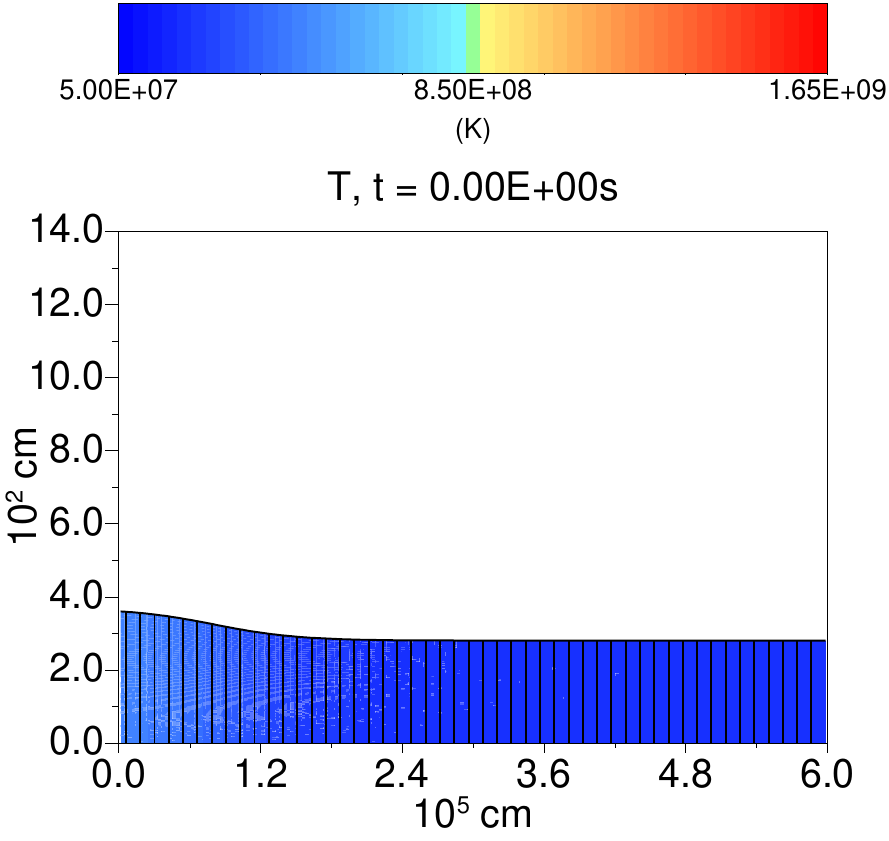}{}{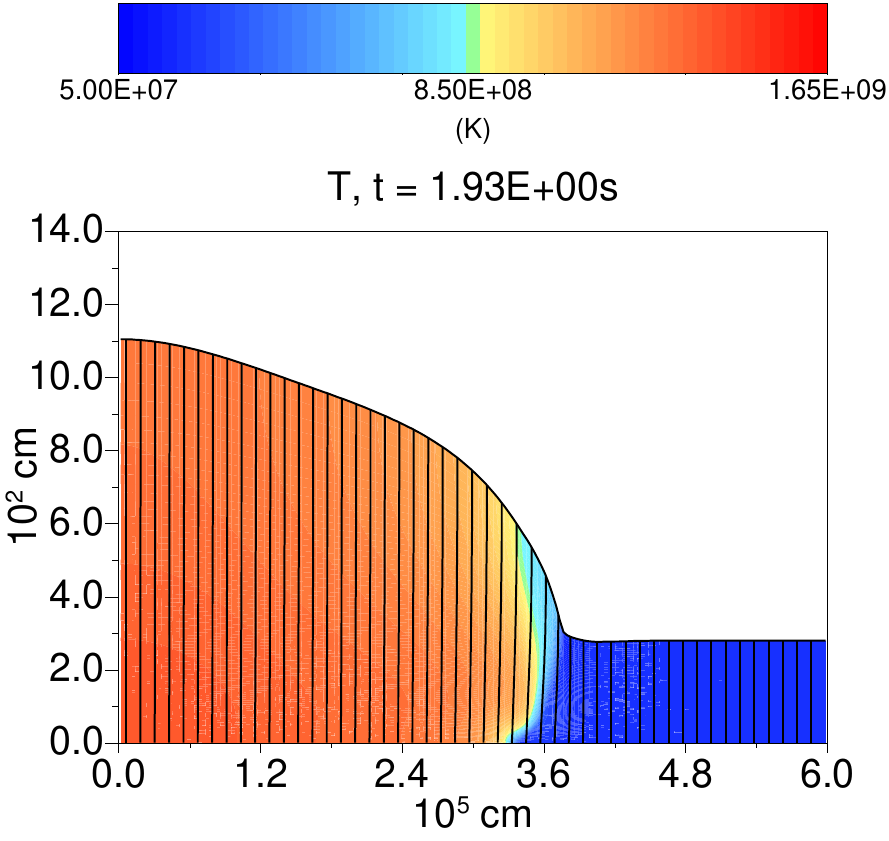}{}
  \doubfig{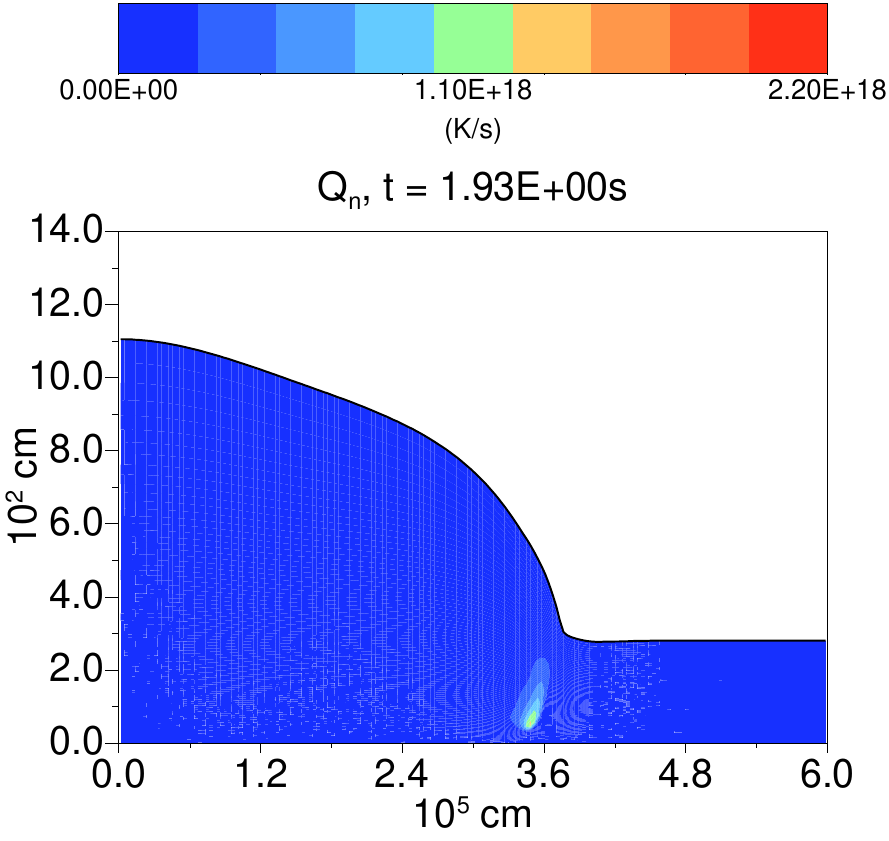}{}{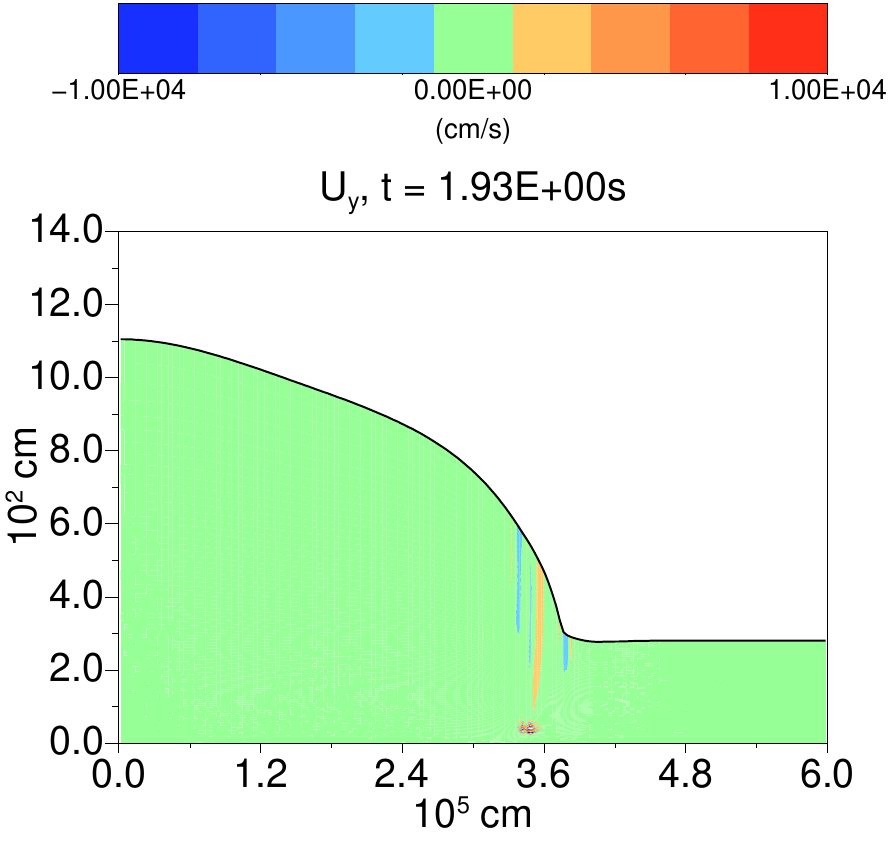}{}
  \caption{\textbf{Upper panels}: initial and late conditions of the
    propagating flame in the case of Run 4, \brep{1}{10} and
    \nrep{450}. The magnetic field lines are superimposed on the
    temperature profile. The magnetic field is strong enough to
    prevent most of the fluid motion. Even when the flame is
    propagating, the magnetic field lines are barely bent.
    \textbf{Lower panels}: nuclear reaction rate (\textbf{left}) and
    velocity in the \y{} direction (\textbf{right}) for the same
    simulation. The absence of $\uy$ shows that the propagation is
    driven by the sole magnetic confinement, since the Coriolis force
    is almost completely suppressed, to the point that a different
    color scale is needed to highlight the small components of the
    velocity.}
  \label{fig:bf4wb}
\end{figure*}

We start with the case of \brep{1}{10}, Run 4. This field strength is
at the upper end of the estimated burster magnetic field
distribution. In this case the initial overpressure at the top of the
temperature perturbation cannot bend the magnetic field lines
significantly. Indeed, the frictional time scale is much faster than
the rotational period (see \tabref{tab:runs}). The initial motion is
quickly suppressed and with no significant \z{} velocity, no
significant Coriolis force appears that can generate the thermal wind.

In \figref{fig:bf4wb} we show the results for Run 4. The top left
panel shows the initial configuration of the magnetic field
superimposed on the temperature profile. This configuration is the same
for all simulations, apart from the value of the seed magnetic field
$\Bo$. On the right, we show the system at a later time: it can be
seen that the field lines have not been bent significantly. The
maximum bending is at the interface between hot and cold fluid, barely
visible at $z \sim 3.6\tent{5}$ cm. We observed the trace particle
sliding down the hot-cold interface when the flame is
approaching. When the flame has passed, they move mostly vertically
along the straightened back field lines. Eventually, they only follow
the expansion of the fluid, not moving anymore with respect to the
fluid itself.

We can compare this simulation to Run 0 (same spin, \nrep{450}, but
without magnetic field) looking at the lower panels of
\figref{fig:bf4wb} and at \figref{fig:bf0wa}. The first significant
feature is that no hurricane structure has developed, as can be seen
by the absence of any strong velocity $\uy$ perpendicular to the plane
in \figref{fig:bf4wb}. Accordingly, the field has been bent in the
\y{} direction only slightly ($\By/\Bx \sim 0.1$ at its maximum). The
only exception is a weak component of $\uy$ right above the flame,
where a component of at maximum $1.5\tent{5}$ cm s$^{-1}$ is seen
briefly oscillating back and forth. We can see also a very weak
component of the velocity at the top of the front, this is associated
with the Coriolis force reaction to the motion of the fluid at the top
of the interface. The second feature is that after a steady state
propagation has been established, the structure of the flame is very
similar to the purely Coriolis confined one, even if the interface
between hot and cold fluid is steeper, in the sense that there is no
turbulence at the front, as opposed to what we see in the other
simulations (see below). If we compare the estimated values for $\tr$
and $\tf$ from \tabref{tab:runs} in the case of Run 4, we can see that
the magnetic coupling should strongly dominate over the Coriolis force
and therefore the length scale for the confinement should also be
smaller than the Rossby radius.

In \citet{\cath} we discussed how the propagation speed scales
proportional to $\propto\lr/H$, since $\lr$ and $H$ set the
inclination of the interface along which conduction can take place and
the shallower the interface (longer $\lr$) the more effective
conduction is. In the general case, $\lr$ should be replaced by the
hot-cold fluid interface horizontal length scale, be it imposed by the
Coriolis force or the magnetic field or a combination of the two
forces. As a confirmation, we find that the flame speed in the present
case is \emph{slower} than in the case without magnetic field since
the flame is confined over a shorter distance. This simulation
corresponds to the case of strong friction discussed in
\secref{sec:results-calc}. The overpressure on the left is acting on
all the layers of fluid ahead, due to strong magnetic vertical
coupling, and the velocity is predominantly in the \z{} direction,
i.e. the horizontal direction of flame propagation in our notation.

\subsubsection{Weak magnetic field ($10^{7}$ {\rm{G}})}
\label{sec:weak}

\begin{figure*}
  \centering
  \doubfig{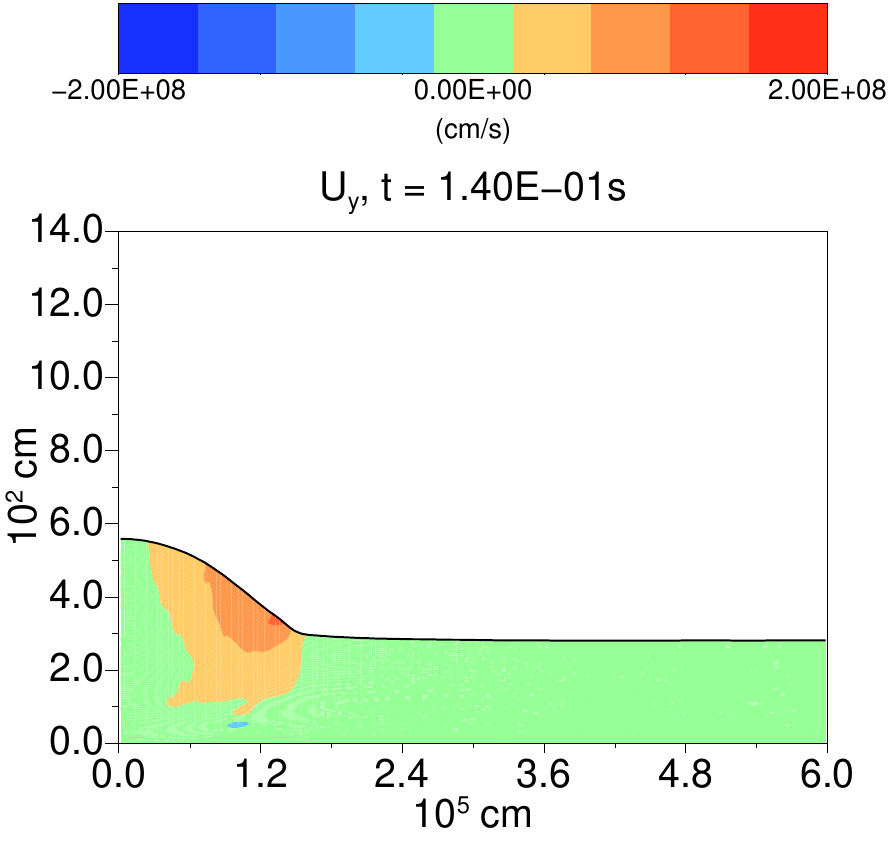}{}{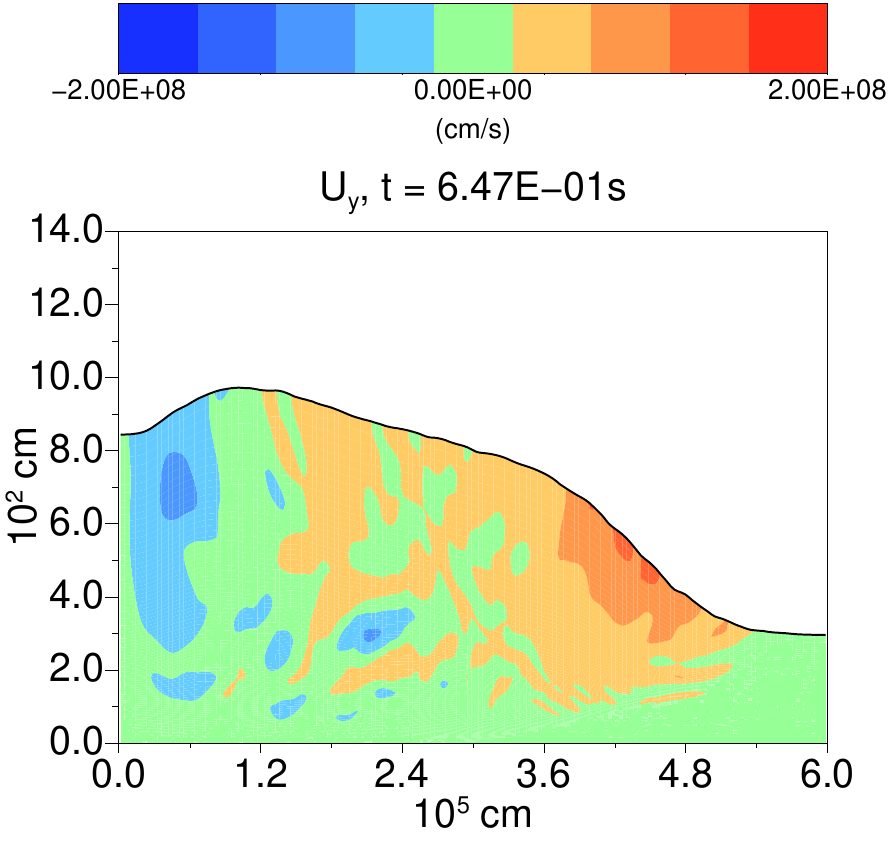}{}
  \doubfig{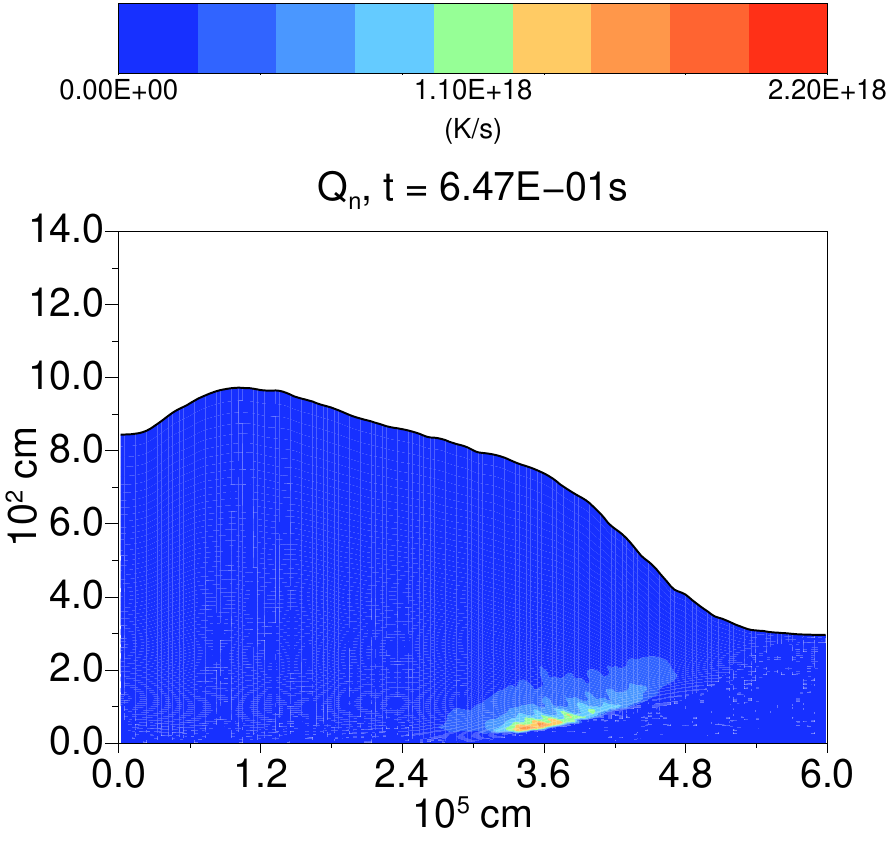}{}{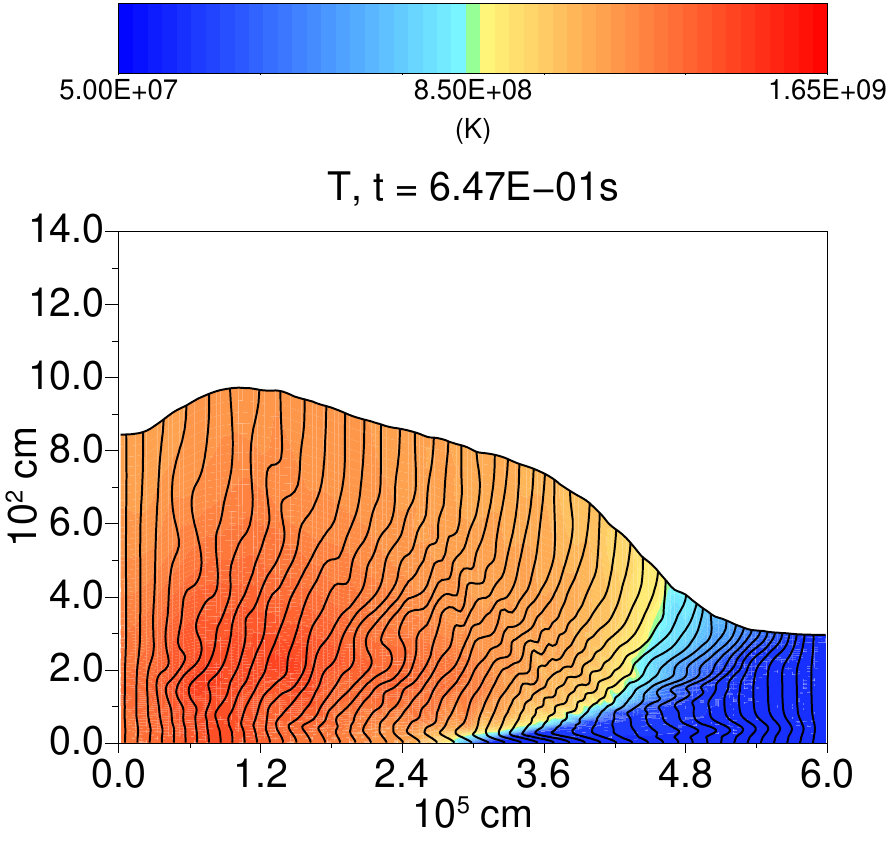}{}
  \caption{Results for Run 1, \brep{1}{7} and
    \nrep{450}. \textbf{Upper panel}: velocity perpendicular to the
    plane $\uy$ at an early (\textbf{left}) and later (\textbf{right})
    stage. Initially the height peak described in the text is not
    present: it develops at $\sim 0.3$ s, when the flame has passed
    the peak horizontal position. \textbf{Lower panel}: burning rate
    (\textbf{left}) and temperature profile (\textbf{right}) in the
    steady state propagation regime. Field lines are overplotted on
    the temperature profile.}
  \label{fig:bf1w}
\end{figure*}

We move now to the other extreme of \brep{1}{7}, Run 1. This
approximately corresponds to the lower end of the magnetic field
distribution for bursters. This case is much more dynamical. As can
be seen in \figref{fig:bf1w} the late evolution of the flame and the
field is strongly influenced by the motion of the fluid. Despite the
fact that we call this case \emph{weak}, it has to be borne in mind
that even here the magnetic field is playing a role in the dynamics.

At the beginning, the magnetic field is not capable of confining the
initial pressure imbalance of the fluid and the fluid spills over
sideways. At \nrep{450}, the Coriolis force is strong enough to
intervene and a hurricane structure begins to develop
(\figref{fig:bf1w}, upper left panel). The fluid motion drags the
magnetic field along and strong horizontal components are
generated. This agrees well with the fact that $\tr \ll \tf$ (see
\tabref{tab:runs}). However, the flame is very different from Run 0
(the case without magnetic field) as can be seen comparing the
lower left panel of \figref{fig:bf1w} to \figref{fig:bf0wa}. In the
case with weak magnetic field, we can see that the hot-cold fluid
interface is much more extended. As a consequence, when we measure the
propagation speed, we find that in the case of \brep{1}{7} the flame
moves faster than in the case of no magnetic field. Note that the
value reported in \tabref{tab:runs} is the result of a \emph{linear}
fit to the horizontal position of maximum burning as a function of
time. In this particular case, we found that the position of the flame
was better fitted by a parabola than a line. This means that the
velocity was accelerating till the flame reached the boundary. The
value we report is therefore an average value.

The reason for the longer interface and the higher speed is the
reaction of the magnetic field which partly obstructs the Coriolis
force. For example, it can be seen that the \y{} component of the
velocity, which characterizes the hurricane structure and ensures the
Coriolis confinement, is smaller in absolute value than in the case
with no magnetic field and is also limited to layers far above the
level where most of the burning takes
place\footnote{\setcounter{foota}{\value{footnote}}On the other hand,
  simulations with \brep{1}{1}, \brep{5}{3} and \brep{1}{6}, have
  velocities $\vf=2.06\tent{5}$ cm s$^{-1}$, $\vf=2.06\tent{5}$ cm
  s$^{-1}$, $\vf=2.14\tent{5}$ cm s$^{-1}$, which are almost identical
  to the one of Run 0. These simulations have a clear hurricane
  structure without the presence of waves along the field lines. In
  these cases, the magnetic field is practically negligible.}. The
magnetic field component $\By$ is maximum at the flame location and
steadily decreases (apart from superimposed oscillations induced by
waves, see below) once the flame has passed. The reduced confinement
is responsible for the longer extent of the interface. This situation
is analogous to the case when friction is beginning to be significant,
but it is still weaker than the Coriolis force. When the flame reaches
the right boundary, the helium fraction is higher than in the case
of no rotation or \emph{strong} magnetic field. This can be easily
understood if we consider that given the higher propagation speed, the
hot fluid burnt for less time.

Finally, we note that after the initial flame has developed,
oscillations set in along the hot-cold fluid interface, near the
flame. Here the baroclinicity is highest and \alf{} waves are excited
and move along the magnetic field lines. The waves propagate also
backwards, at the base of the field lines. These, in turn, excite
other waves along the field lines behind the flame, leading to the
intricate configuration shown in the right, lower panel of
\figref{fig:bf1w}.

There is an additional difference when comparing this case with the
one of \emph{strong} magnetic field or with the one of pure rotation:
after $\sim 0.3$ s a peak forms in the height of the fluid (see
\figref{fig:bf1w}, upper right panel and lower panels). The position
of the peak ($\z \sim 1.0\tent{5}$ cm) does not seem to change
appreciably for the rest of our simulation, and it is close to the
position where our initial temperature perturbation was half of its
maximum value ($\z\sim\z_0$ in \equlab{.} \ref{eq:Tpert}). At this
position the field lines are bent and remain bent for the rest of the
simulation. When the flame reaches the boundary on the right, the
fluid is still burning and eventually the peak disappears.

In order to confirm the correspondence with the initial perturbation,
we ran a further simulation where the initial temperature distribution
was given by \eqr{eq:Tpert}, but with $\z_0=1.5\tent{5}$ cm. We
verified that also in this case a peak develops and it is now at $\z
\sim 1.5\tent{5}$ cm, i.e. where the centre of the new perturbation
is. In both cases the magnetic field keeps memory of the initial
perturbation. The damping factor described in \secref{sec:numerical}
is not the origin of the peak, since simulations with higher vertical
resolution, the same physical parameters, but \emph{without} the
damping showed the same behaviour. The peak was there and even more
pronounced. Inspection of the temperature distribution shows that the
base of the fluid, below the peak, has a higher temperature (see
\figref{fig:bf1w}, lower right panel). It appears that the heat is
deposited in this region, mainly by viscous heating, following the
waves that are localized below the peak. Our viscosity is mostly
artificial and this may produce excessive heating, but if a similar
heating process were proven to be physical, then such a peak could be
the origin of an asymmetry in the emission during the burst rise and
therefore lead to burst oscillations.

\subsubsection{Intermediate strength magnetic field ($10^{8}$ - $10^9$ {\rm{G}})}
\label{sec:intermediate}

\begin{figure*}
  \centering
   \doubfig{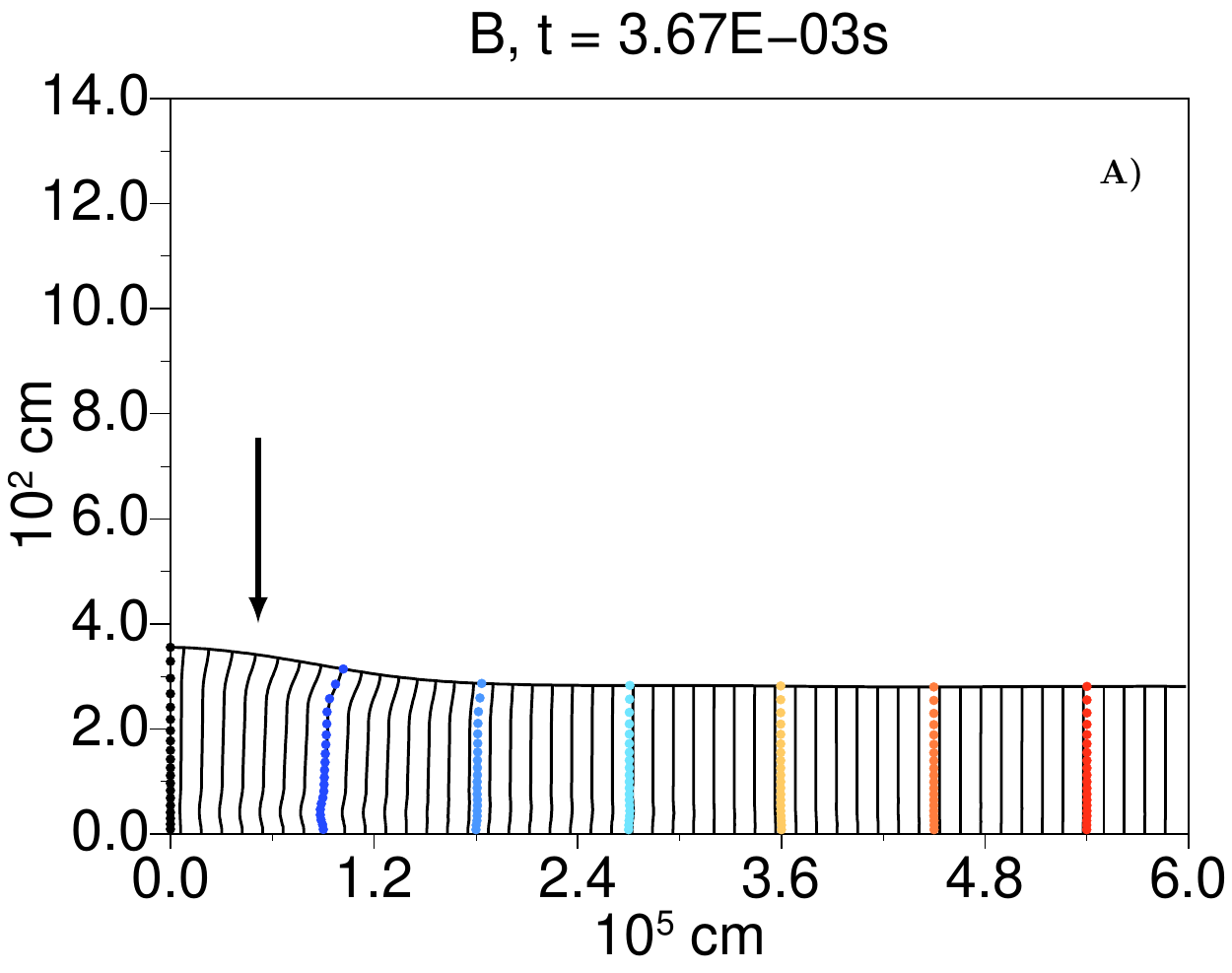}{}{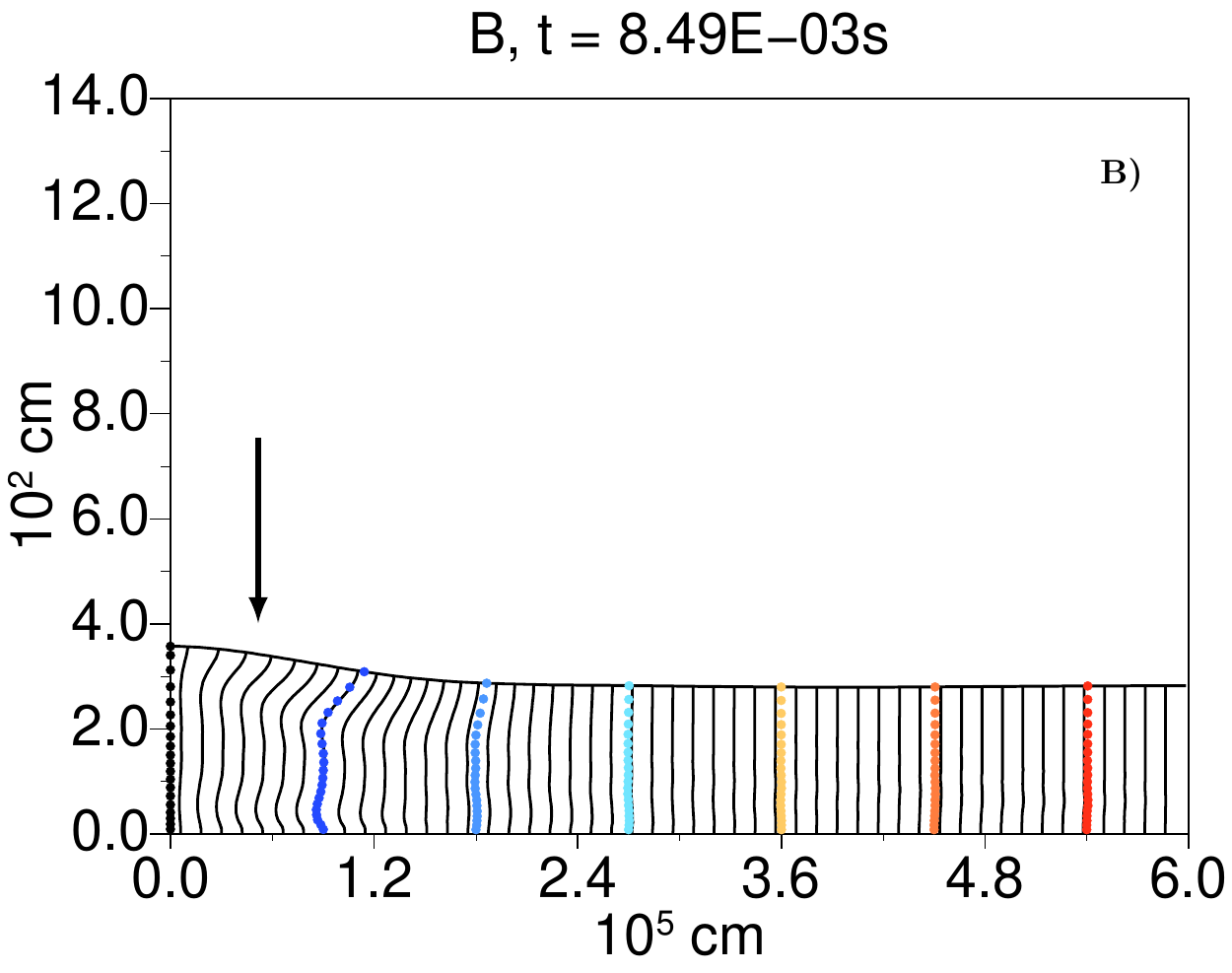}{}
   \doubfig{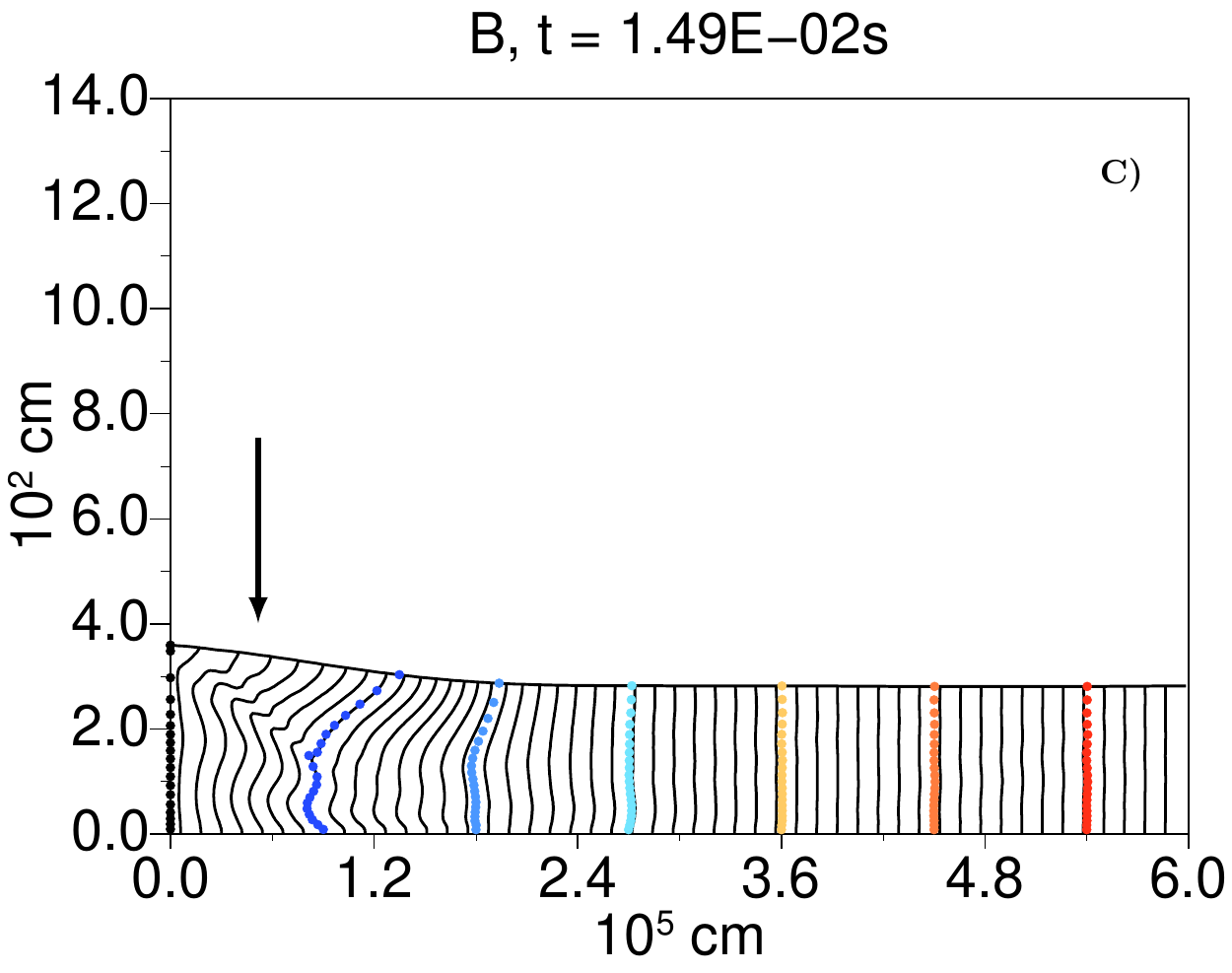}{}{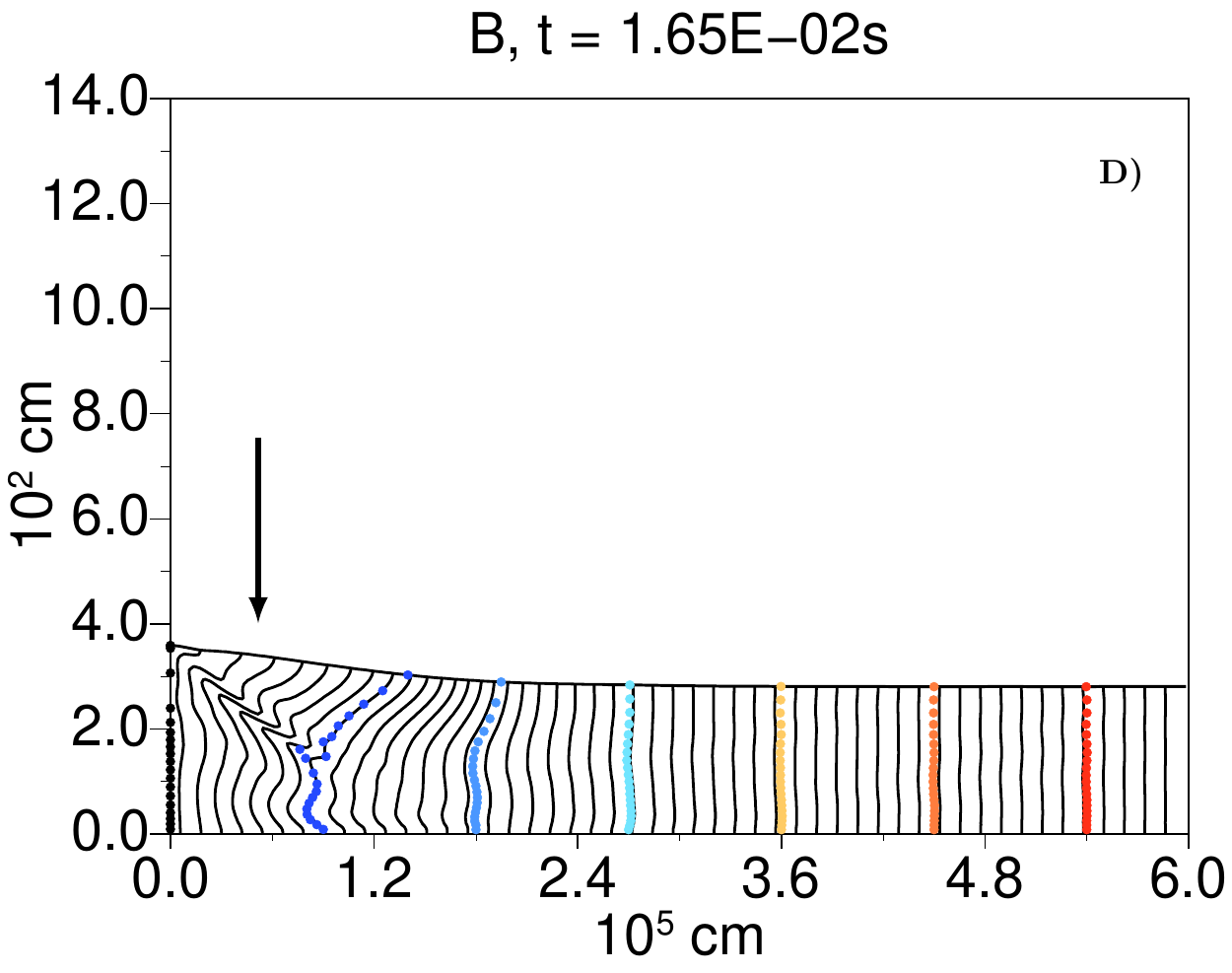}{}
   \doubfig{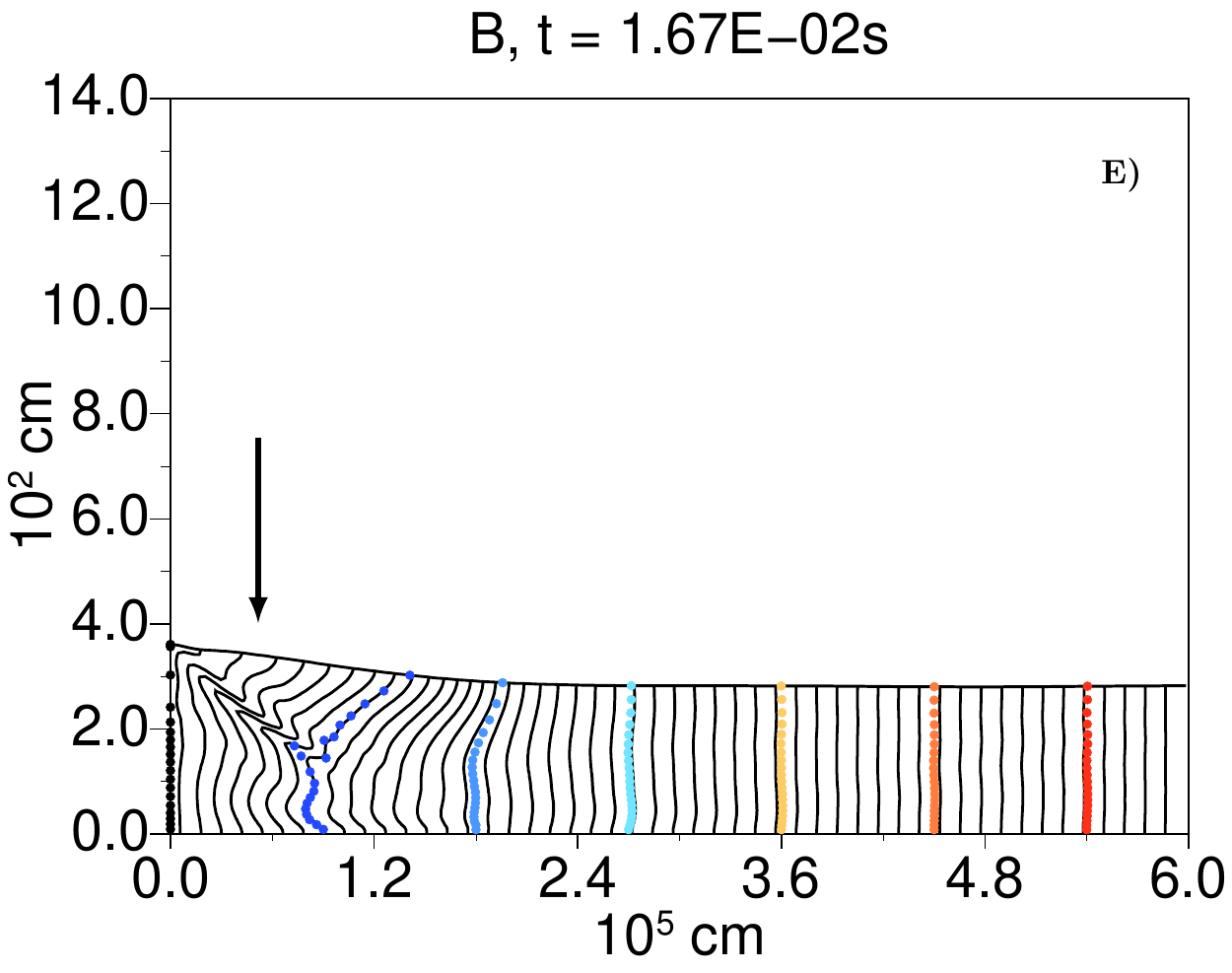}{}{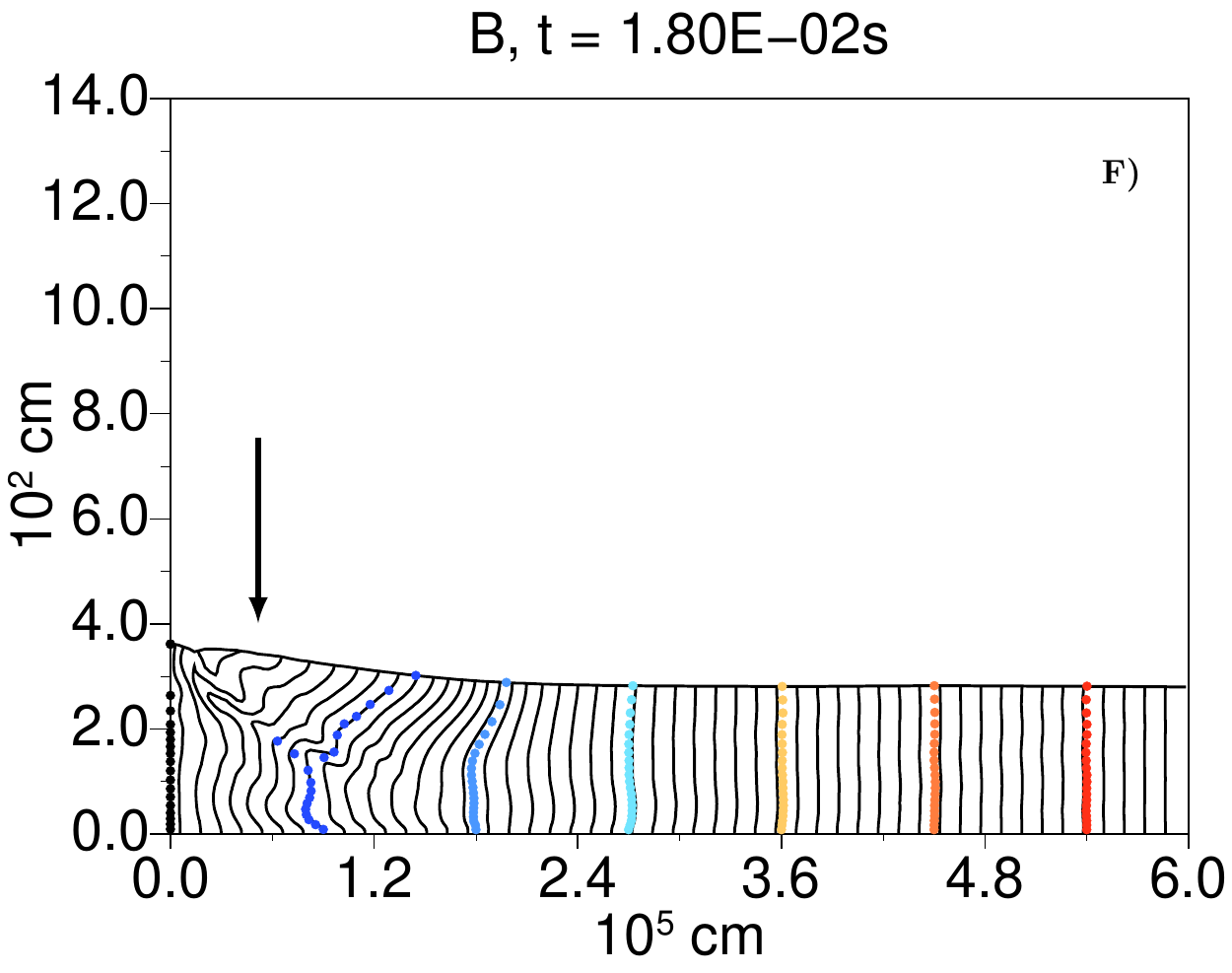}{}
   \caption{Run 2, \nrep{450} and \brep{1}{8}. The initial evolution
     of the thermal structure on the left added at $t = 0$ s to ignite
     the fluid. Shown are the magnetic field lines and tracer
     particles. The color of the particles is related to their initial
     horizontal position. The hot fluid that is trying to spill over
     is eventually blocked and directed downwards resulting in an
     early dilution of the original heat reservoir. The arrows
     indicate the horizontal position of the perturbation described in
     the text.}
  \label{fig:shoot}
\end{figure*}

Finally, we address the case of intermediate strength magnetic field.
The behaviour of Run 2, \brep{1}{8}, is again different from the
previous ones. In this case the initial heat reservoir perturbation
that we impose on the temperature to trigger the burst is partly
diluted during the very early stages of the simulation. The whole
flame propagation is very chaotic.

The partial dilution of the initial heat reservoir perturbation occurs
in the following fashion. As can be seen in \figref{fig:shoot}, the
hot fluid in the upper left corner of the simulation begins to slide
to the right (panels A and B). The field lines follow the fluid,
bending near the top at an approximate height of $ \sim 2\tent{2} -
3\tent{2}$ cm. Then, the curvature of the knee of the bent field lines
increases and secondary, alternating, knees form below (panels B and
C). The configuration becomes unstable around the upper knee (panel D)
with the hot and cold fluid mixing: the field lines reconnect and the
hot fluid moves right- and downwards, while the cold fluid goes up-
and leftwards. This triggers a cascade of fluid rolls that propagate
right- and downwards (panels D, E and F). Analogously, later in the
simulation, we see a similar perturbation affecting the field
lines. The perturbation propagates like a wave triggered where the
weight of the fluid that is moving near the top of the simulation
bends the field lines and these form a knee. The wave propagates from
the formation of the knee down- and rightwards across the field lines:
see the configuration of the field line around $\z\sim1.8\tent{5}$ cm
in \figref{fig:bf2wext1} or the field line just after
$z\sim2.4\tent{5}$ cm at $t\sim3.89\tent{-2}$ s in
\figref{fig:bf2wext2}. This perturbation propagates towards the right
of the simulation (see for example the field lines around
$\z\sim4.8\tent{5}$ cm in the second column of \figref{fig:bf2wext3}),
while additional waves are excited along the field lines that have
been already hit by the front of the perturbation. Note that at $t\sim
1.11\tent{-1}$ s another front of perturbation is now reaching $\z
\sim 4.8\tent{5}$ cm (\figref{fig:bf2wext3}), at $t\sim1.3\tent{-1}$ s
it will be at $\z\sim 5.4\tent{5}$ cm and it will reach the right
boundary at $t\sim 1.4 \tent{-1}$ s. In order to find in which range
of magnetic field strength the same \emph{initial} bending of the
field lines that leads to the beginning of the perturbation takes
place, we ran a few more simulations and found that the same
phenomenon is seen down to $\Bo\sim3\tent{7}$ G, but not anymore at
$\Bo\sim5\tent{8}$ G. Note that this range encompasses the critical
value $\bar{\Bo}$ of \eqr{BZcr}.

The partial dilution of the heat reservoir did not cause the flame to
die, and the burning began right at the start of the simulation, but
the flame is different from the well confined flame of the other cases
we have treated so far. A pseudo steady state propagation is
established: from the main burning region, hot fluid tongues are
launched up- and rightwards, these eventually turn downwards and
leftwards when they reach regions where magnetic field lines have been
compressed together. Secondary flames keep igniting where the front of
the perturbation brings hotter fluid to regions of higher density. The
new flame will propagate \emph{backwards} till it joins the main
burning region. First examples of these secondary burning blobs can be
seen in \figref{fig:bf2wext2}. Clearer examples can be seen in
\figref{fig:bf2wext3}, since the burning is by now more intense also
in the new flames, where the association of the secondary ignition
sites with compressed field lines is more clear. The fluid keeps
rolling between regions where the field lines have been compressed,
mixing hot and cold fluid. An example can be spotted noting the
deformed shape of the field lines between $\z\sim 2.4\tent{5}$ cm and
$\z\sim3.0\tent{5}$ cm at $t\sim 1.17\tent{-1}$ s in the lowest right
panel of \figref{fig:bf2wext3} or in the right, top panel of
\figref{fig:bf2wext7bf3w} between $\z\sim 4.2\tent{5}$ cm and
$\z\sim4.8\tent{5}$ cm.

These rolling eddies appear where the fluid is more baroclinic and
driven by the baroclinic version of buoyancy. In this sense they are a
form of convection \citep{book-1987-Pedlo}. The fluid follows the
compressed field lines on its left when going upwards, and the motion
that follows after the fluid is stopped by the denser field lines
ahead is reminiscent of the behaviour of the Parker instability
\citep{art-1966-park-1}. This is understandable when considering the
configuration of the field lines in this simulation. The field has
been stretched and has developed a strong horizontal component. When
the fluid at the top is stopped, it rests on these stretched lines,
which are naturally subject to the instability. The fluid slides down
and backwards along the inclination of the lines.  When it stops, this
fluid has higher entropy than the fluid above it and consequently is
pushed vertically upwards again, generating the rolling motion between
the regions of denser field lines. The continuous ignition of the
fluid repeats this process until the right boundary is reached and the
cycle cannot proceed anymore.

The flame proceeds forward very fast, $\vf \sim 1.6\tent{6}$ cm
s$^{-1}$, in this fashion, until by $t\sim 3\tent{-1}$ s it reaches
the boundary, where the field lines are being compressed. Finally, the
fluid temperature reaches $T\sim 10^{9}$ K everywhere and by $t \sim
5.98\tent{-1}$ s the fluid is expanding vertically
\emph{approximately} homogeneously and the field lines are
straightening up. \emph{We note that these propagation timescales are
  in very good agreement with the observed rise times of observed
  bursts.}

Given the very fast initial phase where the flame reaches the extent
of the boundary while the temperature does not rise above $10^9$ K,
most of the fluid is still unburnt and it will be burned during the
subsequent homogeneous expansion of the whole layer. Comparing with
the case of no magnetic field (Run 0) we find that during the whole
propagation the hurricane structure is present, but the perpendicular
velocity $\uy$ is of smaller amplitude, (the maximum absolute
amplitude is roughly half the one of the case with \brep{1}{7}, Run 1,
or of rotation only, Run 0). This is again in agreement with the fact
that the flame velocity for Run 2 is faster than for Runs 0 and 1. The
smaller the perpendicular velocity, the smaller the Coriolis force and
the greater the propagation speed. In the left panel of
\figref{fig:bf2-3wUy} we show the profile of the perpendicular
velocity $\uy$, where the thermal wind is perturbed by the reaction of
the Coriolis force to the horizontal $\z$ motion of the rolling eddies
and the \alf{} waves excited along the field lines. Finally we note
that during the flame propagation after $t\sim8.46\tent{-2}$ s
(\figref{fig:bf2wext2}) there is a temperature (and height) peak that
is moving with the flame. The thermal wind is positive (out of the
plane) at the flame front and negative (into the plane) behind. This
propagation would appear as an expanding ring on the surface of the
neutron star, at least until the effects of curvature become
important. This is different from what we saw in the case of
\emph{weak} magnetic field, since there the peak was remaining at the
same horizontal position, while the burning region expands, but also
this configuration may lead to burst oscillations during the burst
rise.

We also ran a simulation with \brep{1}{9} (Run 3, see
\figref{fig:bf2wext7bf3w} lower panels). Despite the fact that the
simulation with \brep{1}{8} is much more chaotic than this one and
that no dilution of the initial heat reservoir takes place when
\brep{1}{9}, these two cases share some similarities. The steady state
configuration of Run 3 has a burning region which is more homogeneous
than the one of Run 2, but still it shows secondary burning bubbles
and rolling eddies at the front (visible at $3.6\tent{5} \rm{cm}
\lesssim \z \lesssim 4.2 \tent{5} \rm{cm}$ in the lower panels of
\figref{fig:bf2wext7bf3w}). This simulation bears similarities also
with the case of \emph{strong} field, since the spilling of the fluid
is more controlled than in the cases of lower magnetic field. However, the
front is much more elongated than in that case, and flame propagation
is consequently faster. The flame is faster also than in the case of
no magnetic field, Run 0, but \emph{slower} than in the case of
\brep{1}{7}, Run 1, and \brep{1}{8}, Run 2. The perpendicular velocity
$\uy$ is even smaller than in the case of \brep{1}{8} (Run 2), which is
to be expected since the magnetic stress back-reaction increases with
the seed magnetic field intensity. Moreover, the $\uy$ velocity field
at the front is oscillating and no real thermal wind is present (right
panel of \figref{fig:bf2-3wUy}). This is again very similar to what we
see in the case of \emph{strong} field, so that the case of
\brep{1}{9} should be regarded as a transition between the
\emph{intermediate} and \emph{strong} field regimes. This behaviour is
consistent with the fact that in this case $\tf<\tr$
(\tabref{tab:runs}) which implies that at this point the magnetic
coupling has become stronger than the Coriolis force.

Run 2, \brep{1}{8}, is the configuration that yields the fastest
propagation speed among our configurations. This corresponds to a
situation where the frictional coupling is most effective in
increasing the speed of the flame and the speed value is near its
peak. Indeed, \brep{1}{8} is very close to our estimate of the
critical field $\bar\Bo$. \citetalias{\slu} discussed how in this case
the flame propagates so fast that the ignition would be almost
horizontally uniform: that is indeed what we observe. Increasing the
field strength even more increases the coupling further, and the flame
velocity will begin to decrease. The speed is expected to decrease
because the pressure imbalance at the top has to push both the top and
lower layers that are strongly coupled by the magnetic field. This is
what we see happening in the case of Run 3, \brep{1}{9} and more
drastically in the case of Run 4, \brep{1}{10}.

\begin{figure*}
  \centering
  \doubfig{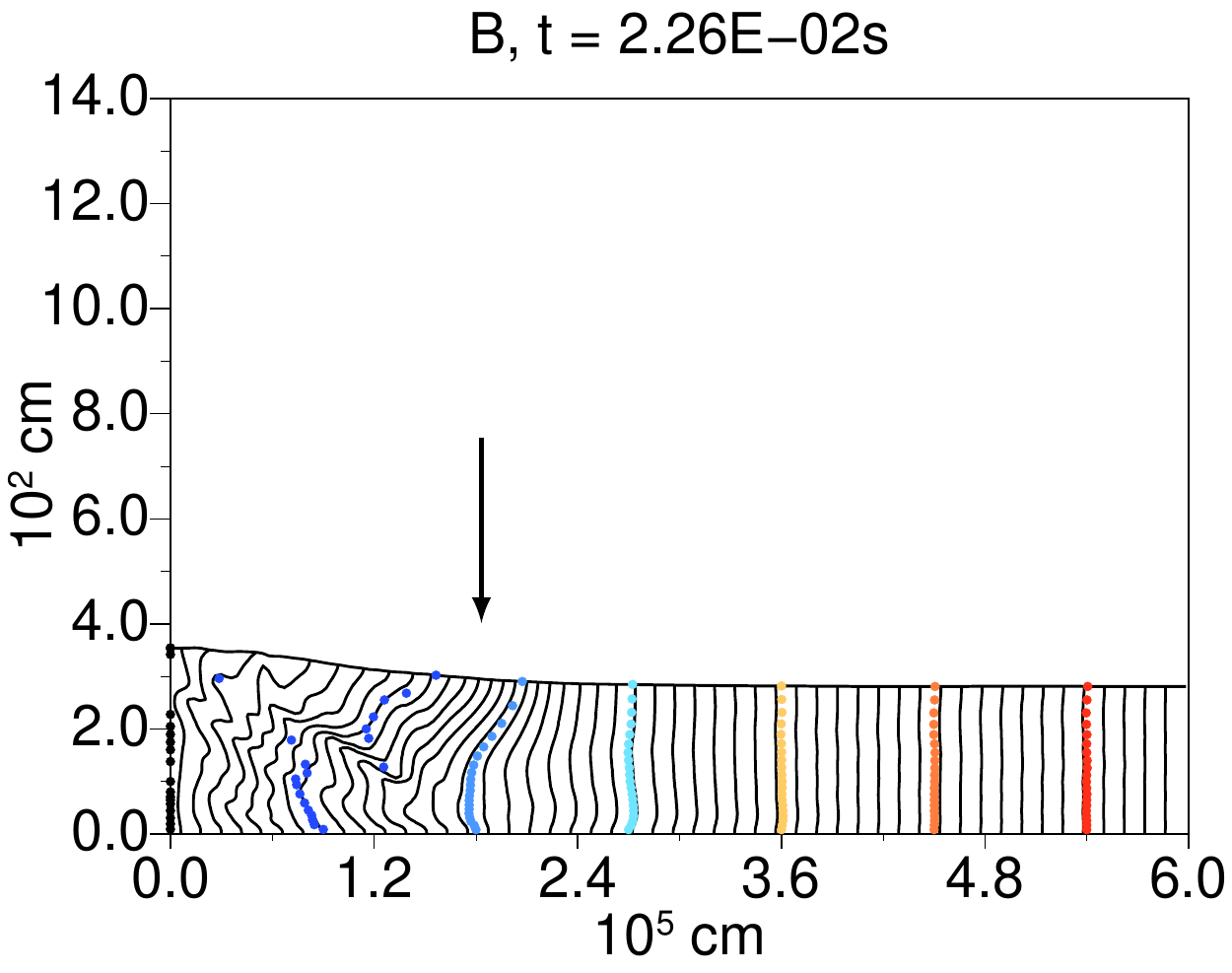}{}{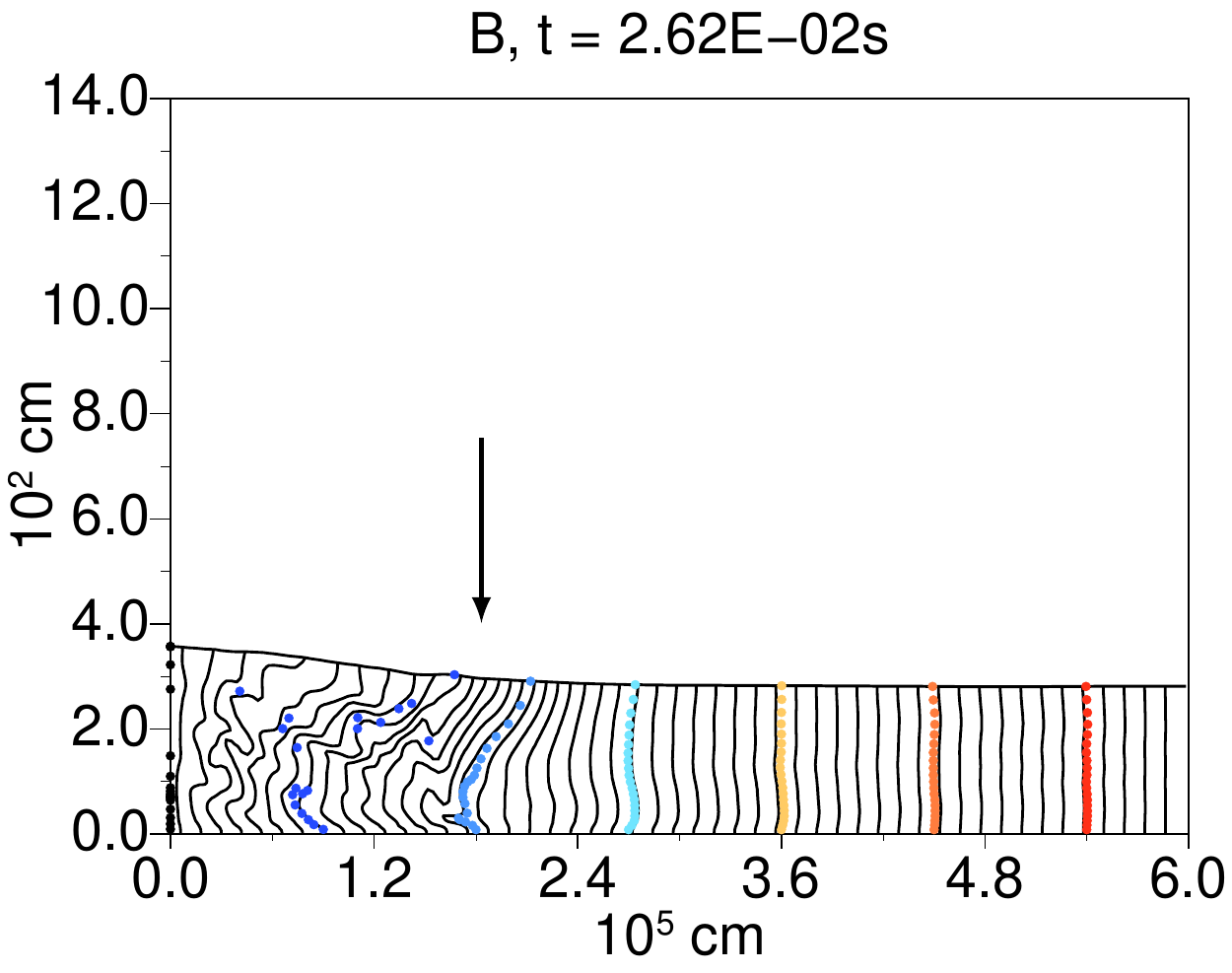}{}
  \doubfig{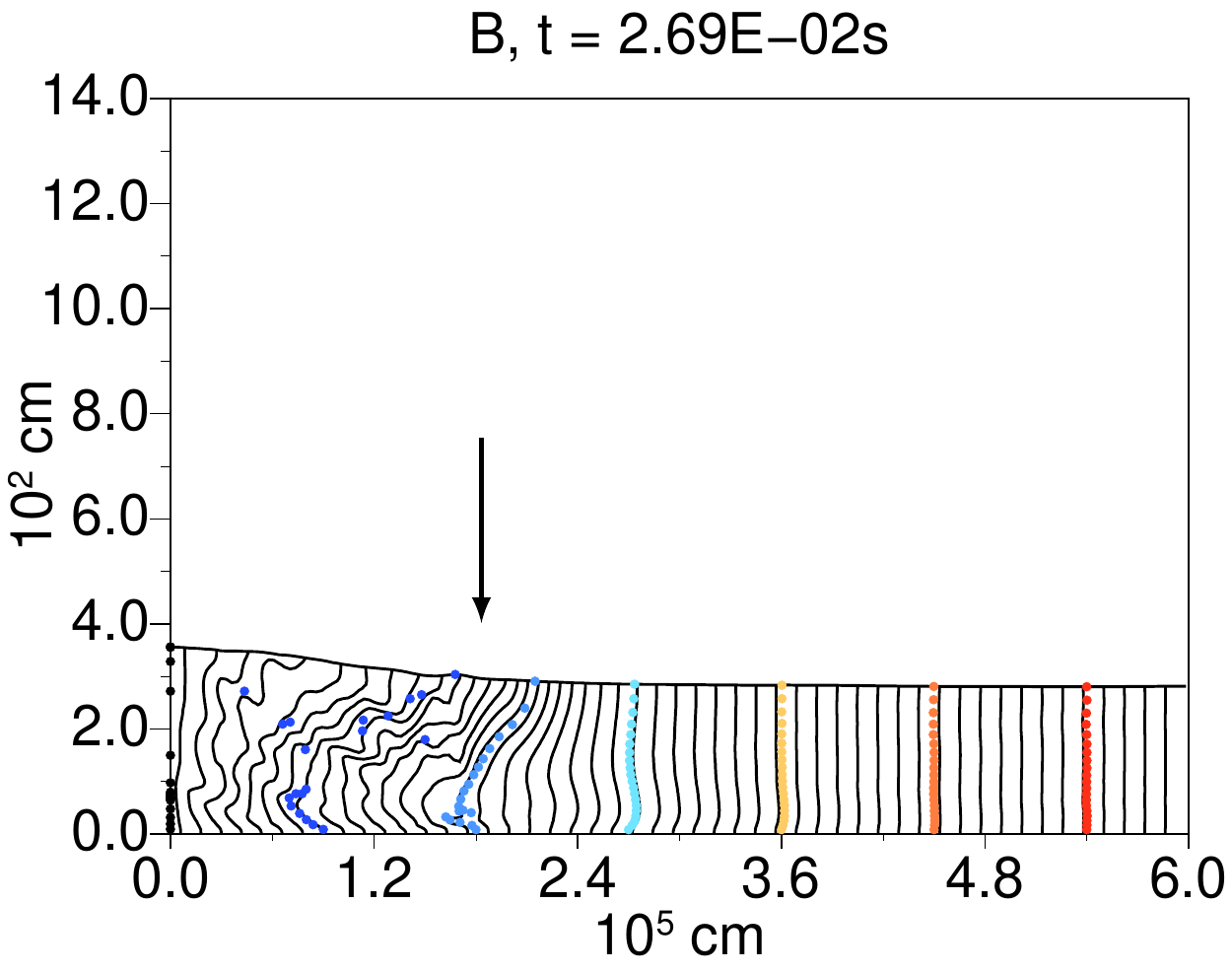}{}{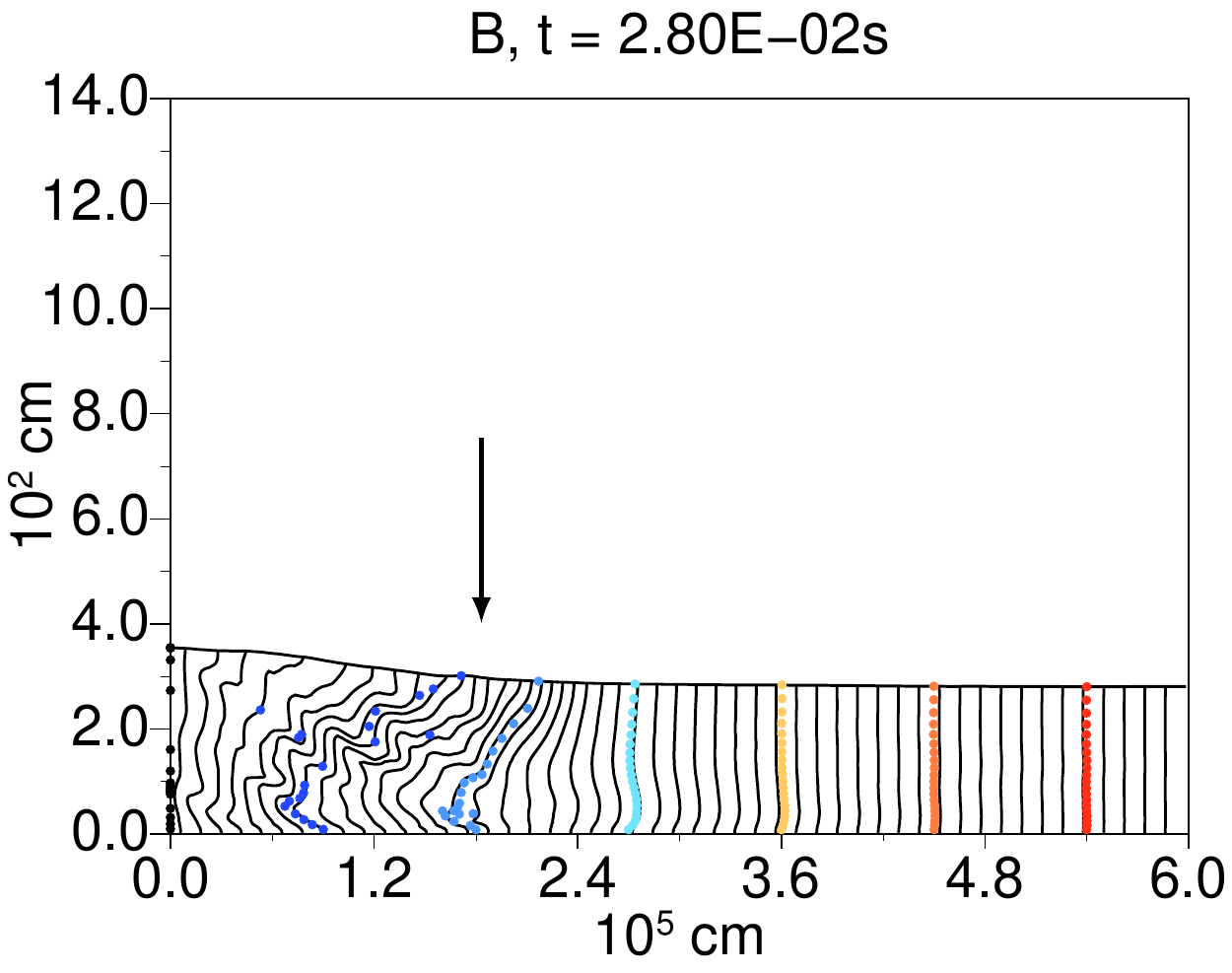}{}
  \caption{Continuation of the early evolution of Run 2 from
    \figref{fig:shoot}, \brep{1}{8} \nrep{450}. Shown are the magnetic
    field lines with tracer particles superimposed. The arrows
    indicate the horizontal position of the later evolution of the
    perturbation described in the text.}
  \label{fig:bf2wext1}
\end{figure*}

\begin{figure*}
  \centering
  \doubfig{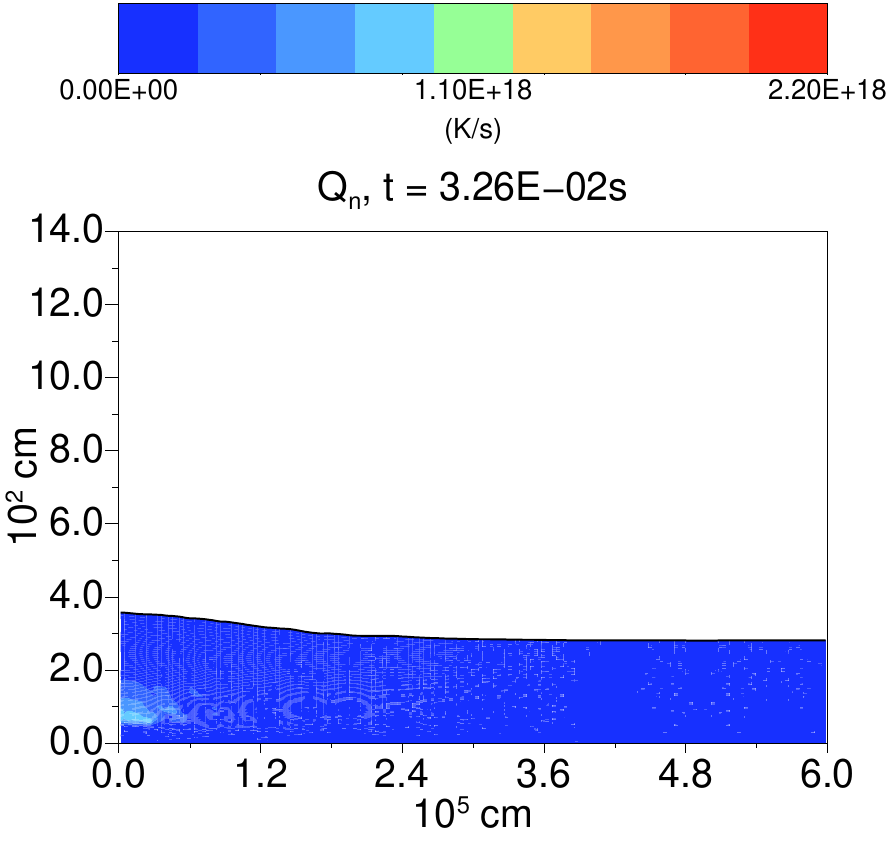}{}{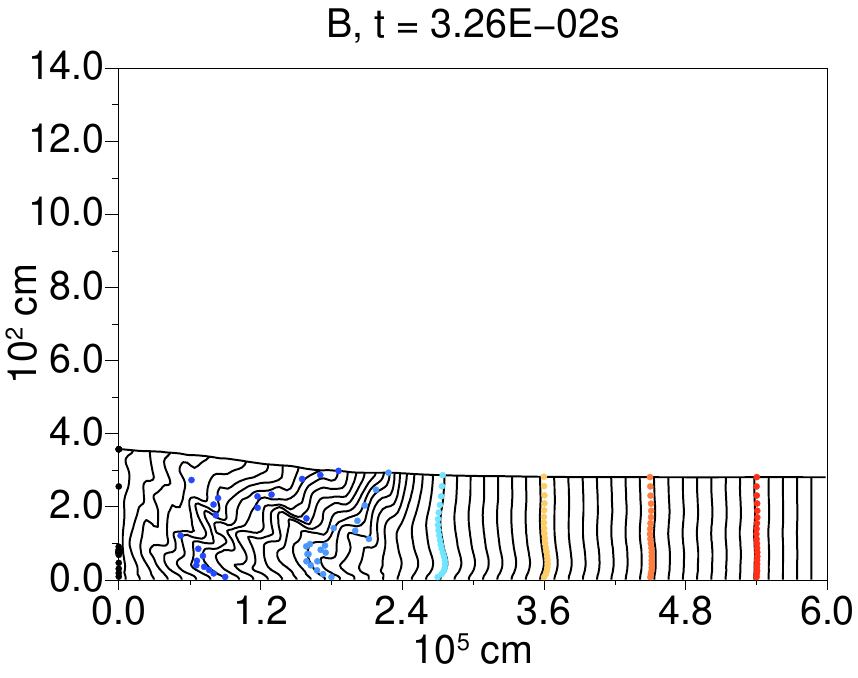}{}
  \doubfig{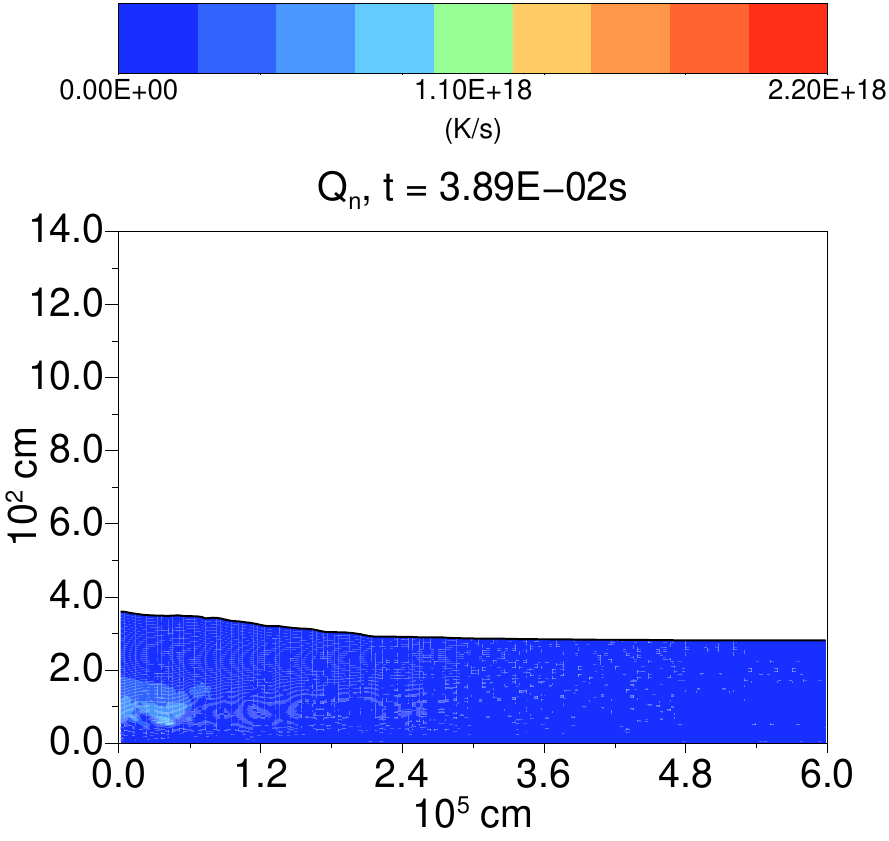}{}{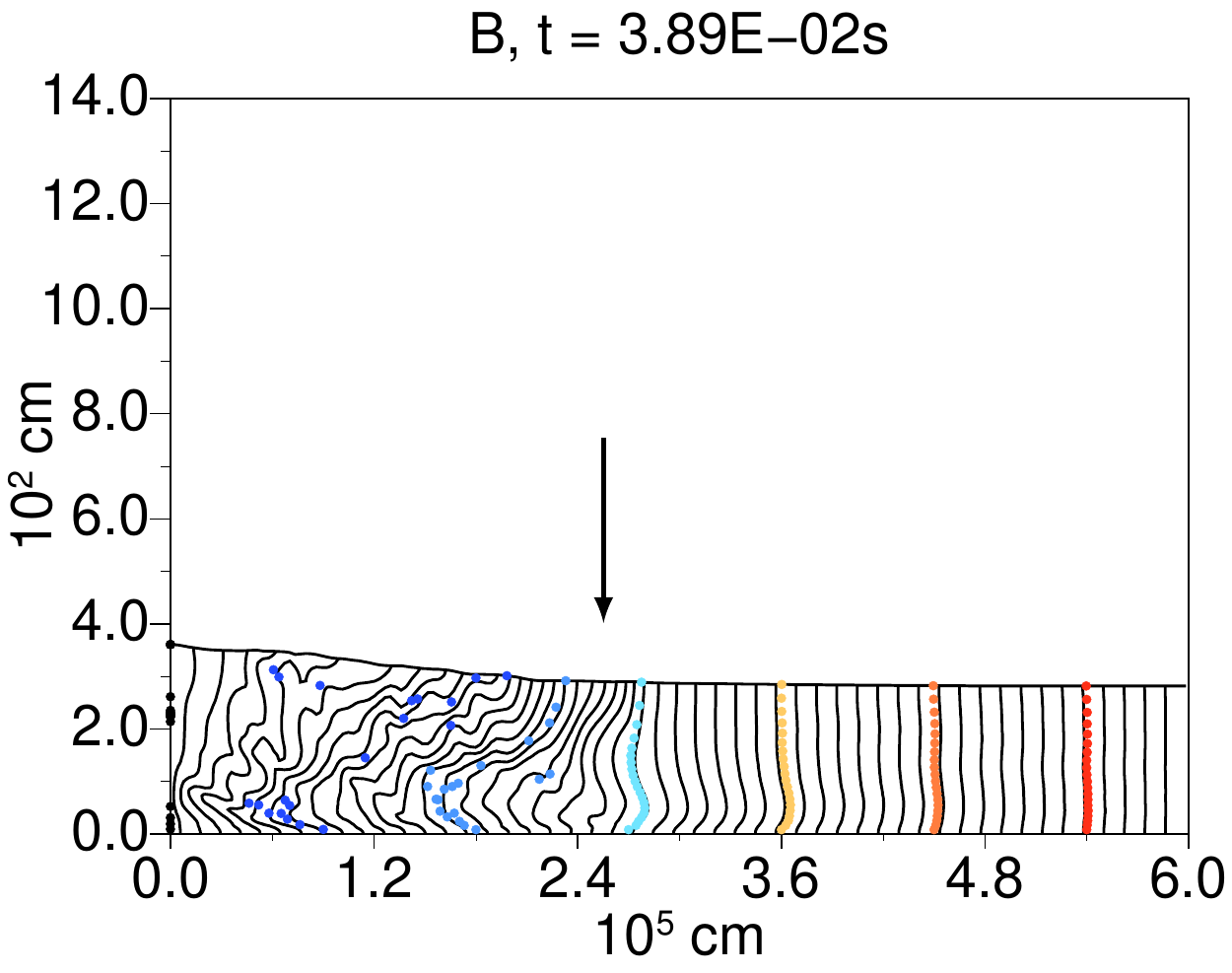}{}
  \doubfig{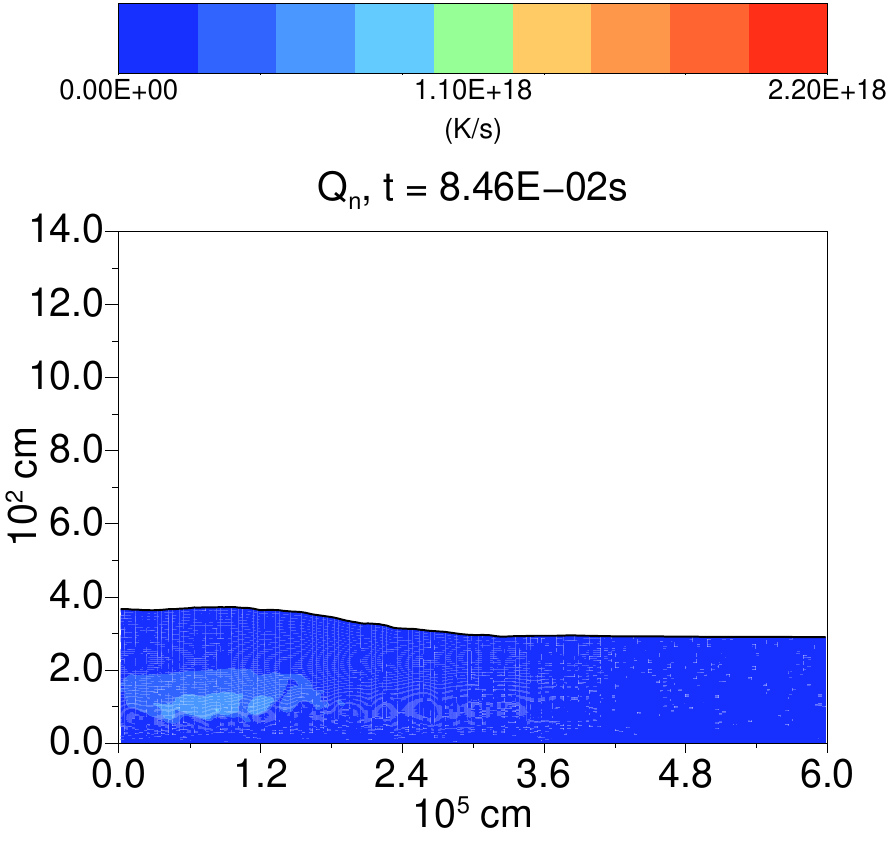}{}{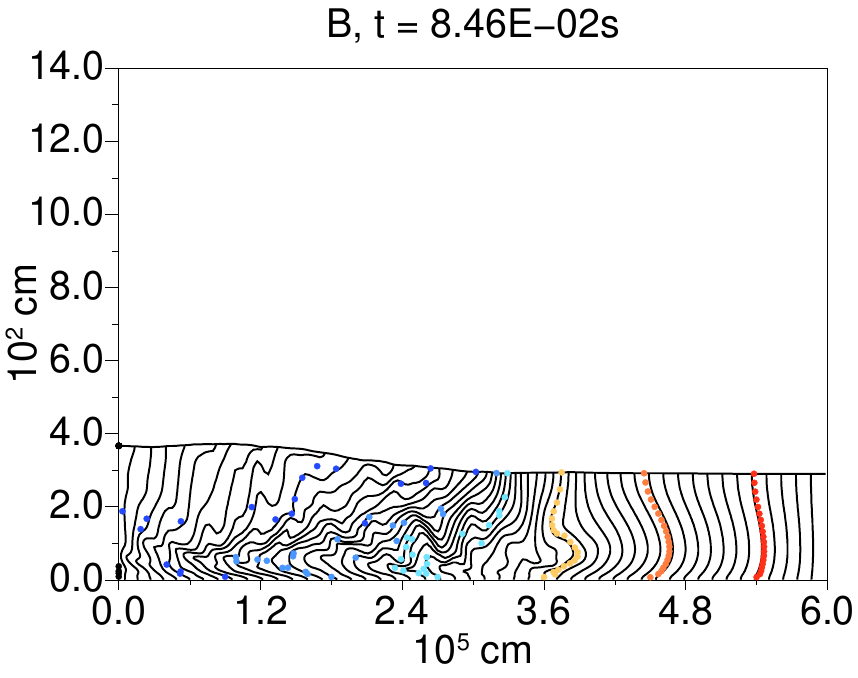}{}
  \caption{Continuation of the early evolution of Run 2 from
    \figref{fig:bf2wext1}, \brep{1}{8} \nrep{450}. Shown are the
    burning rate (\textbf{left}) and the magnetic field lines
    (\textbf{right}), with tracer particles superimposed. The arrow
    indicates the horizontal position of the perturbation described in
    the text at later time.}
  \label{fig:bf2wext2}
\end{figure*}

\begin{figure*}
  \centering
  \doubfig{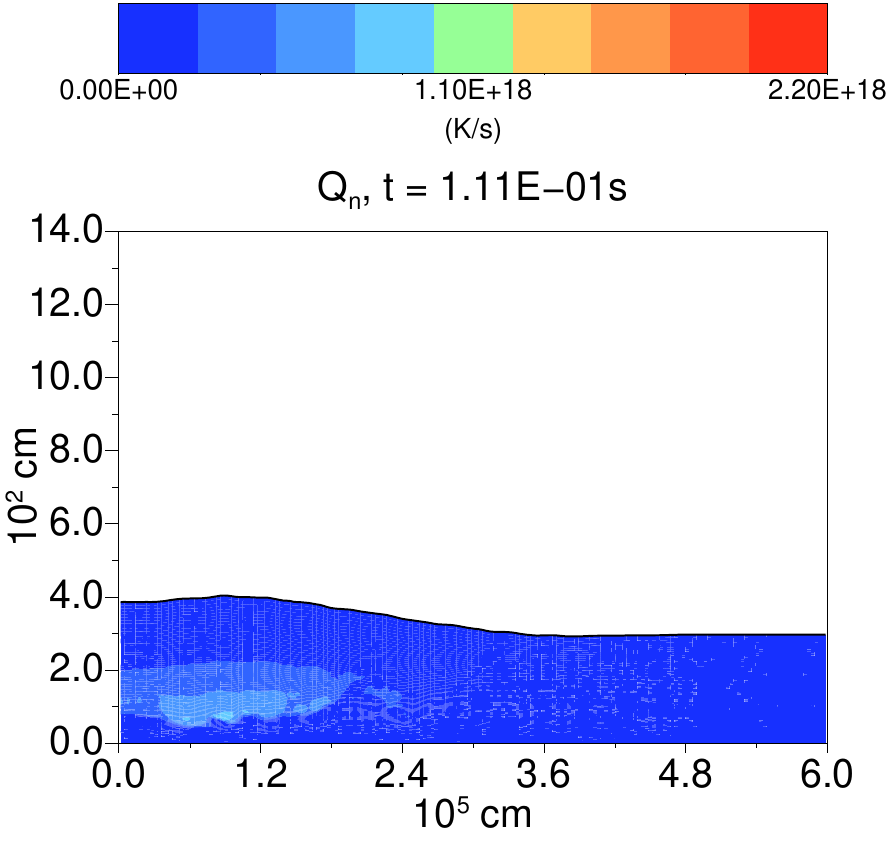}{}{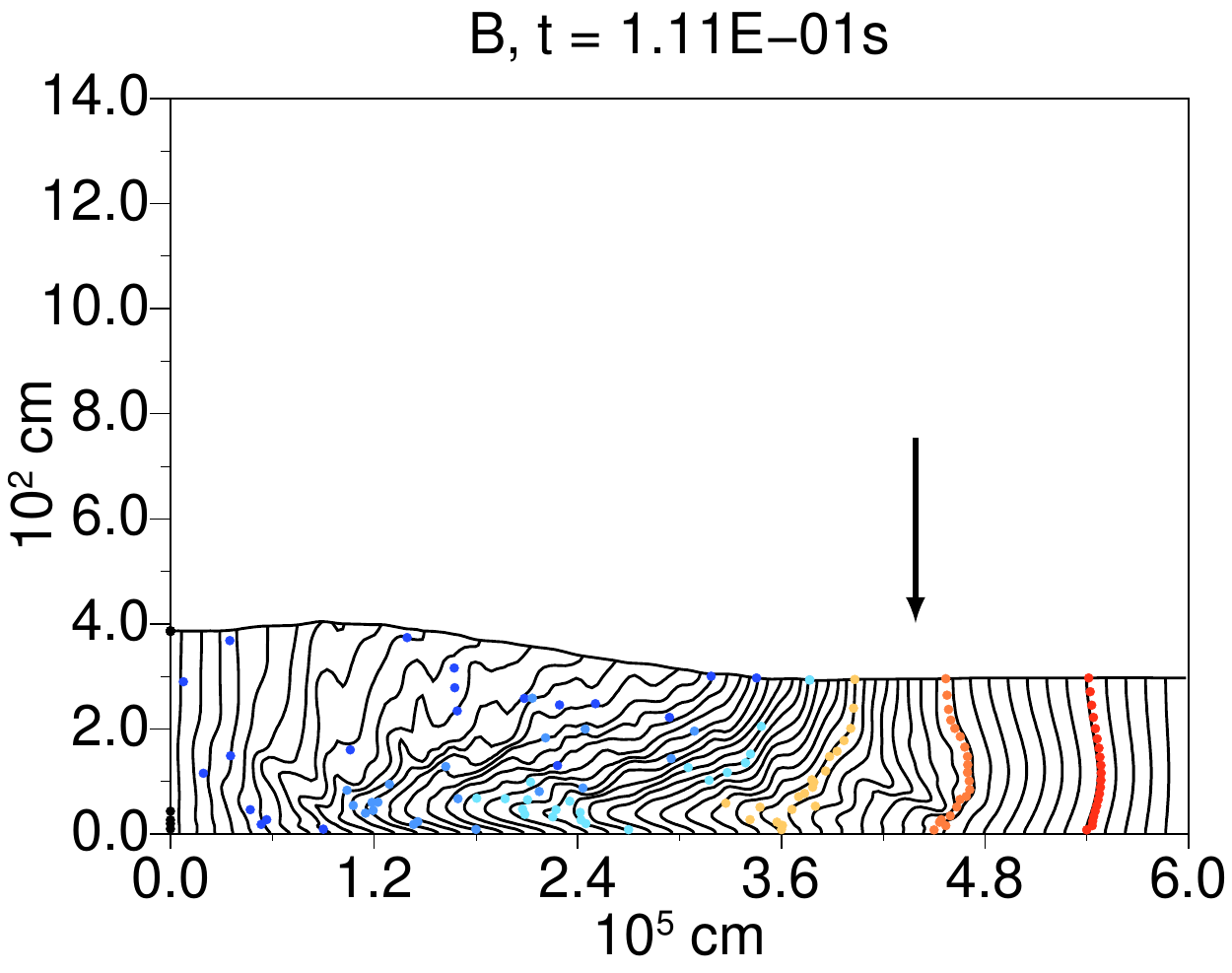}{}
  \doubfig{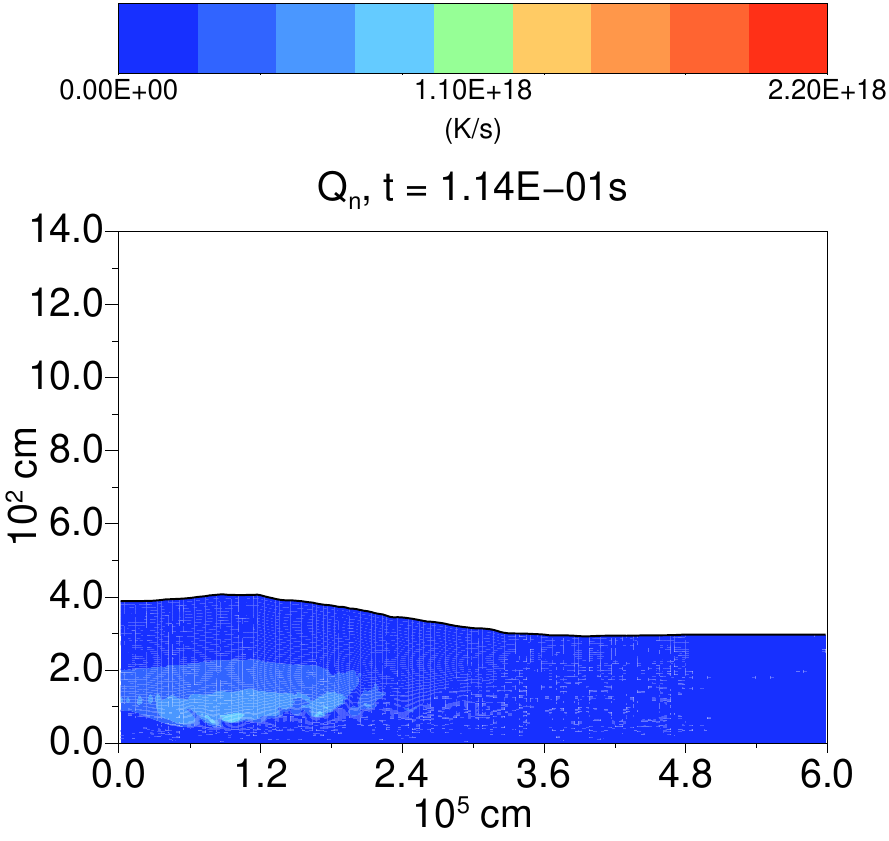}{}{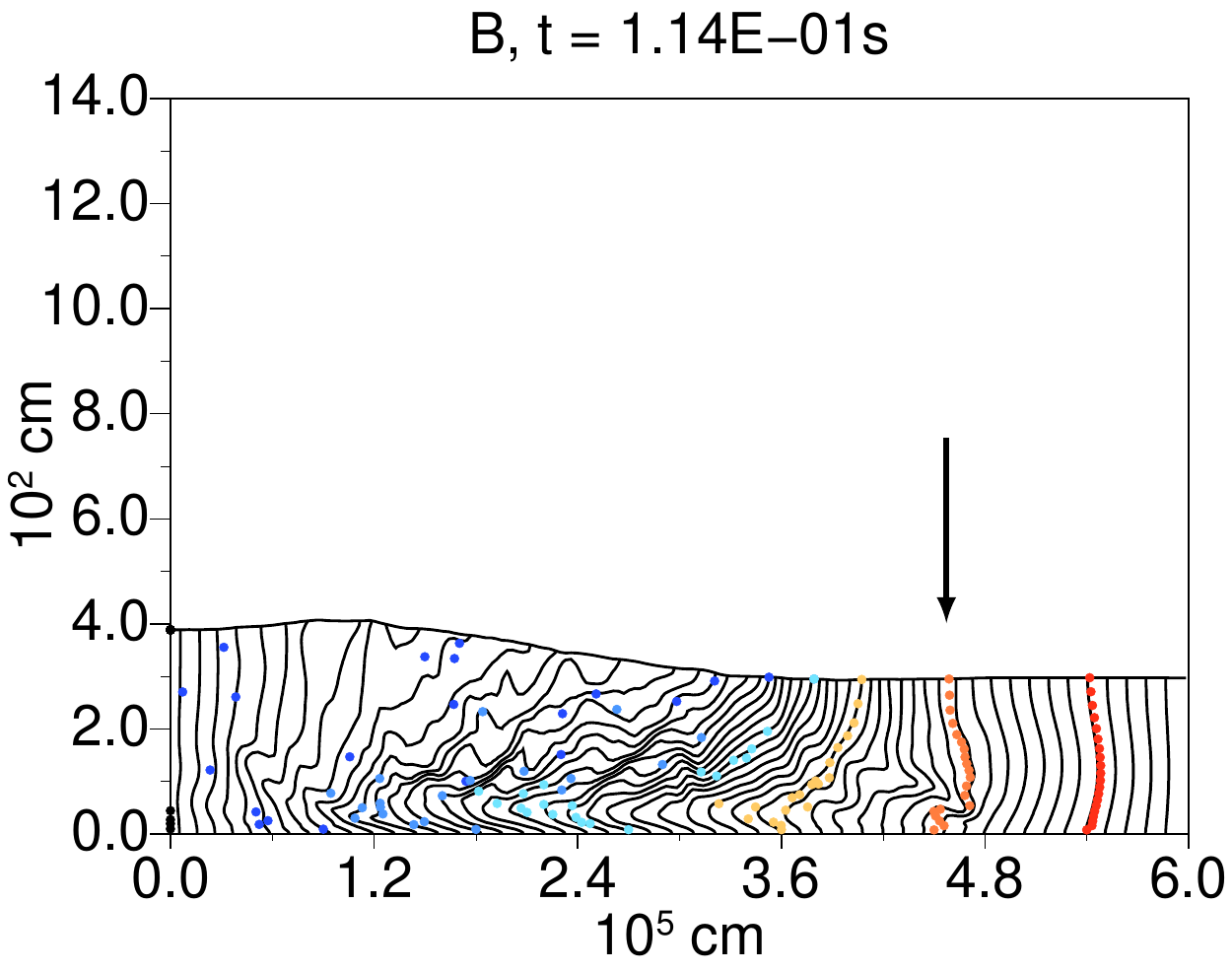}{}
  \doubfig{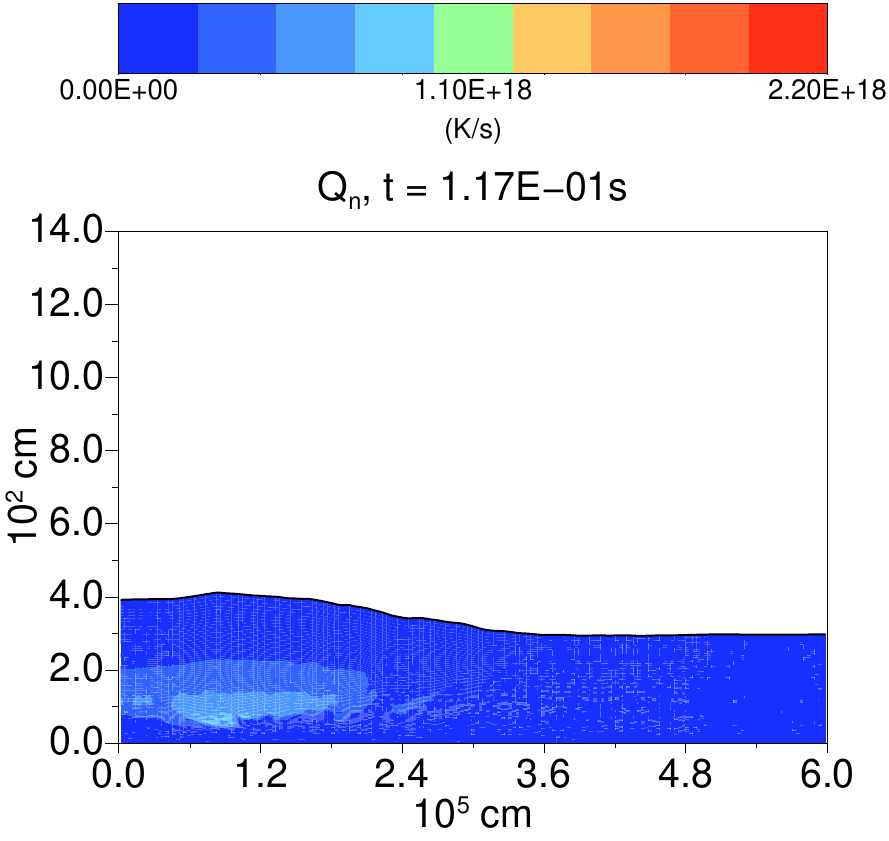}{}{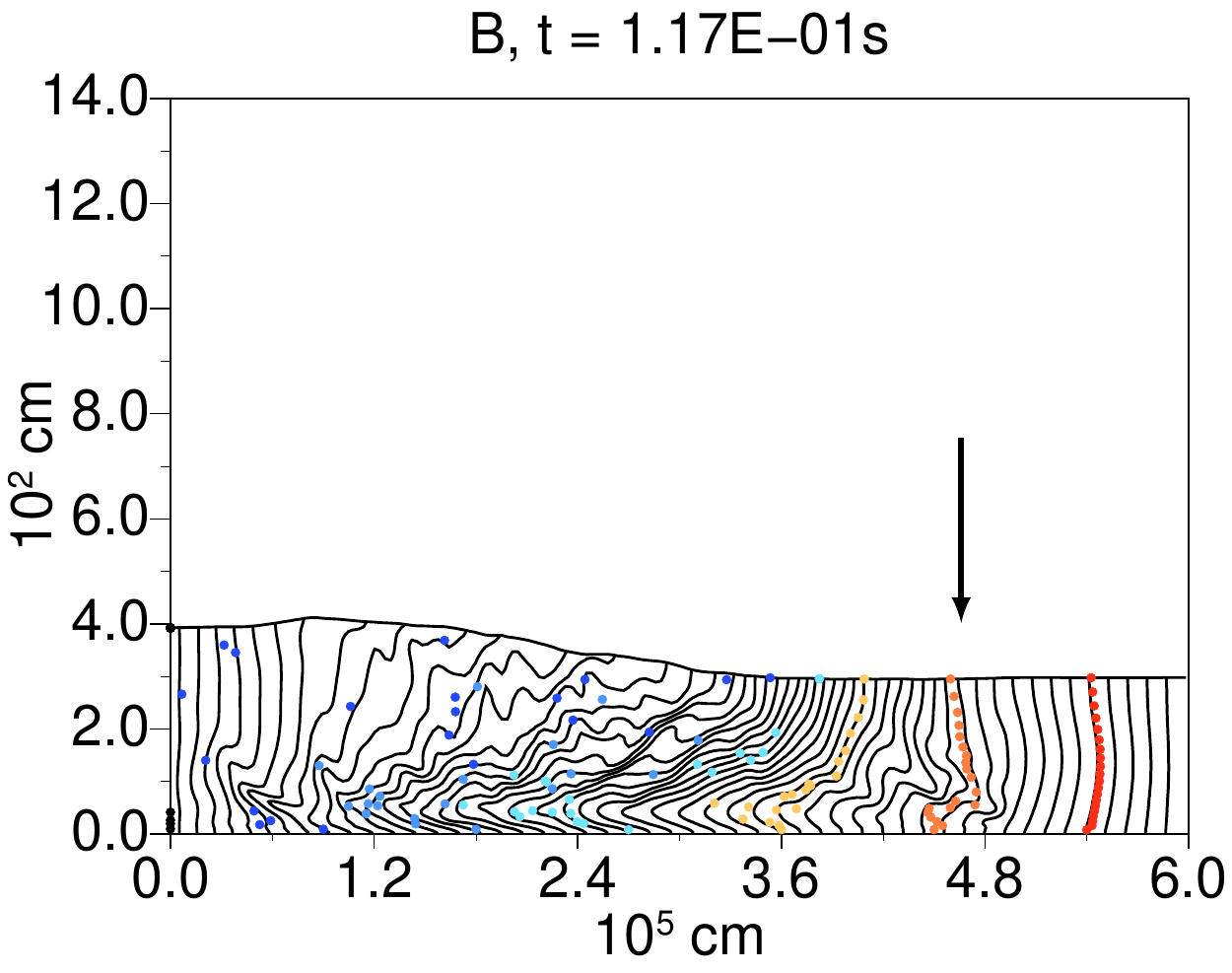}{}
  \caption{Continuation of the early evolution of Run 2 from
    \figref{fig:bf2wext2}, \brep{1}{8} \nrep{450}. Shown are the
    burning rate (\textbf{left}) and the magnetic field lines
    (\textbf{right}), with tracer particles superimposed. The arrows
    indicate the horizontal position of the perturbation described in
    the text at later time.}
  \label{fig:bf2wext3}
\end{figure*}

\begin{figure*}
  \centering
  \doubfig{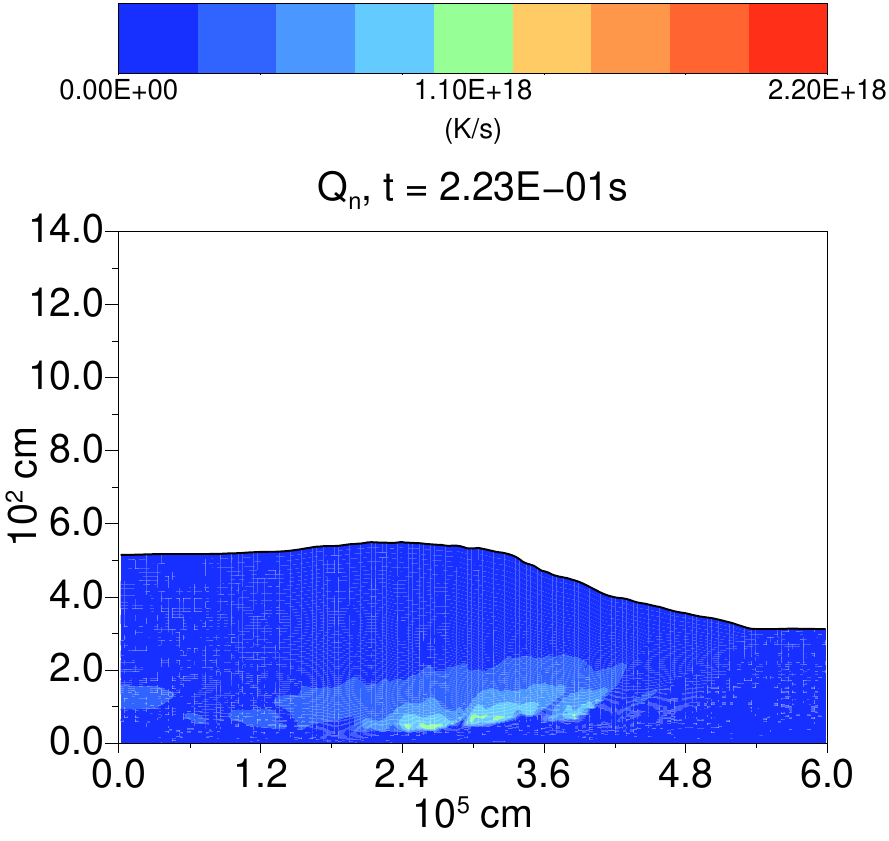}{}{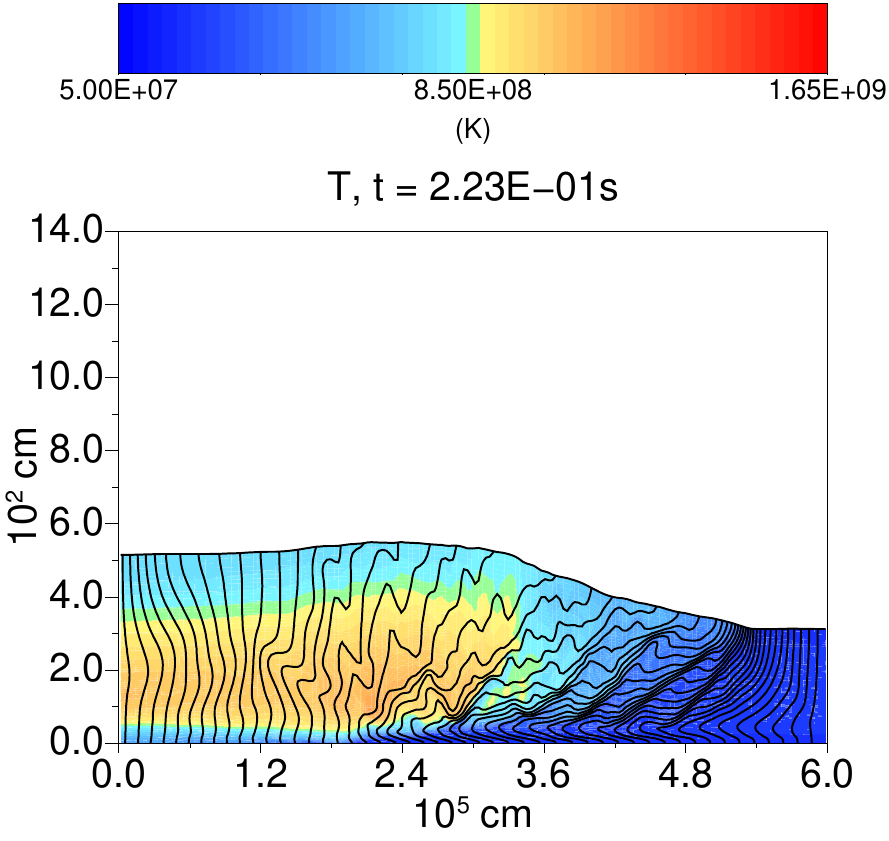}{}
  \doubfig{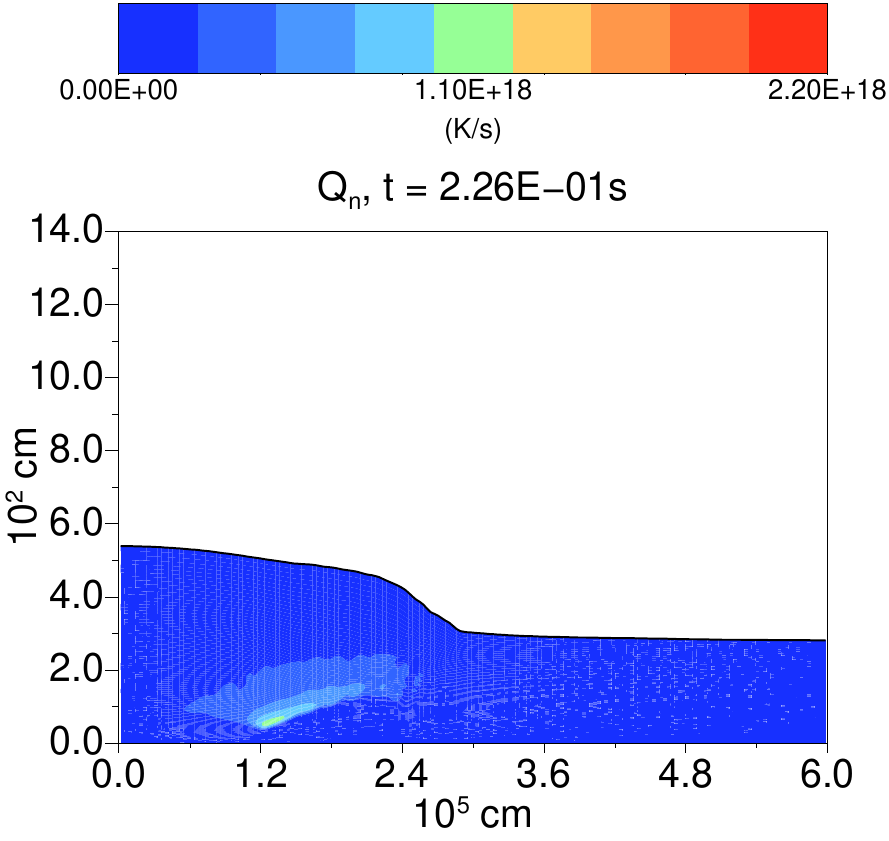}{}{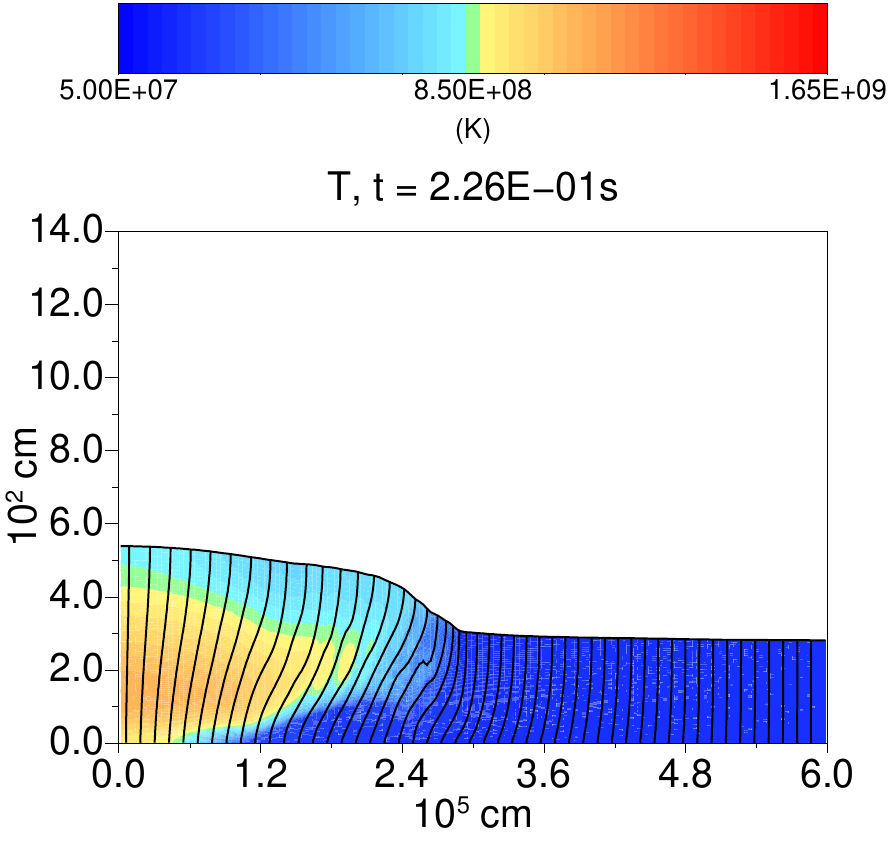}{}
  \caption{The burning rate (\textbf{left}) and the temperature
    profile (\textbf{right}), with magnetic field lines superimposed
    of the later stages of the flame propagation for Run 2,
    \brep{1}{8} \nrep{450}, (\textbf{upper panels}) and Run 3,
    \brep{1}{9} \nrep{450} (\textbf{lower panels}).}
  \label{fig:bf2wext7bf3w}
\end{figure*}

\begin{figure*}
  \centering
   \doubfig{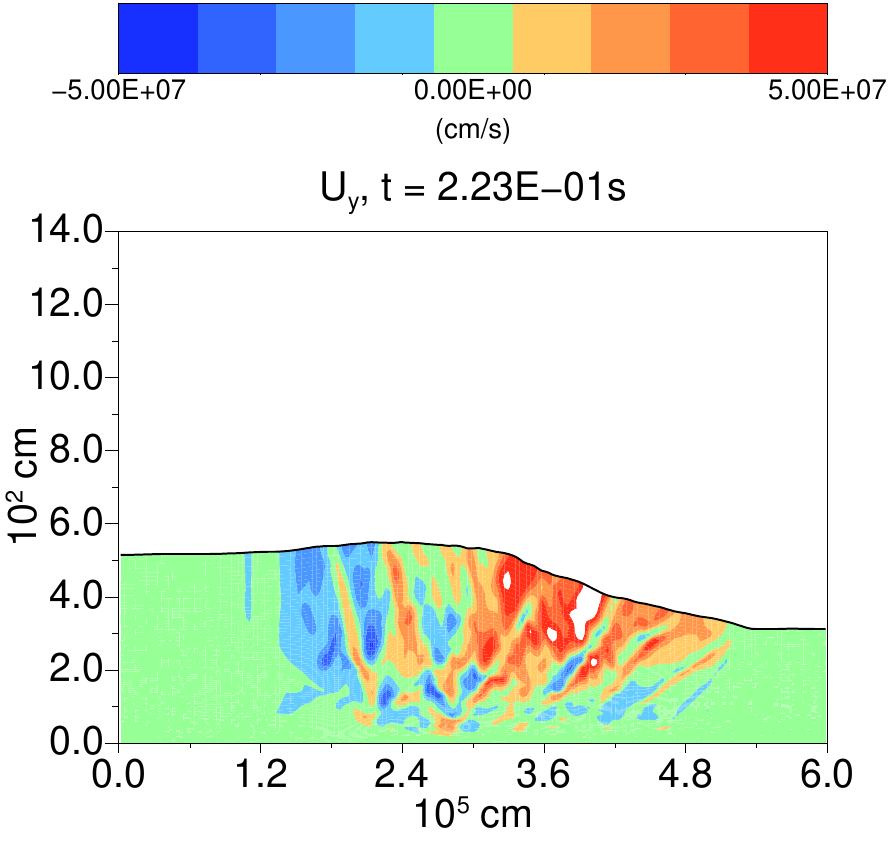}{}{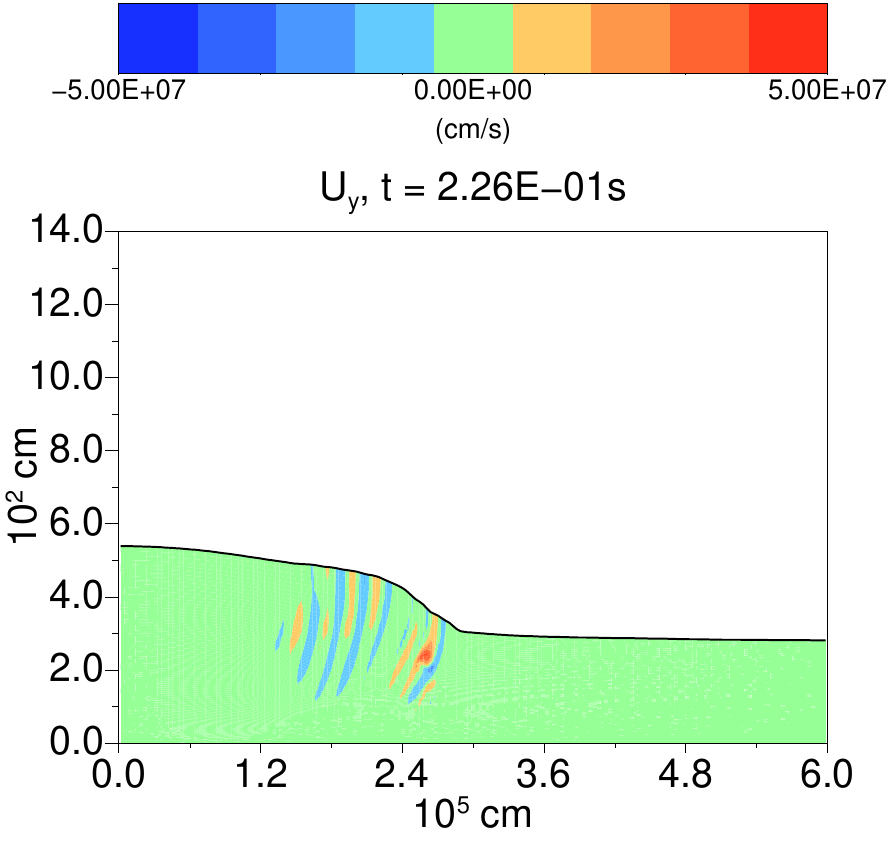}{}
  \caption{The profile of the perpendicular velocity $\uy$ in the case
    of \brep{1}{8} (Run 2, \textbf{left}) and \brep{1}{9} (Run 3,
    \textbf{right}) and \nrep{450}. In the first case the thermal wind
    structure is still present, however disturbed by the reaction of
    the Coriolis force to the rolling eddy motions and by the \alf{}
    waves along the field lines, while in the second case no hurricane
    has developed. Note that the color scale is different from the
    ones of \figrefs{fig:bf0wa}, \ref{fig:bf4wb} and
    \ref{fig:bf1w}. The white color patches correspond to velocity
    values above $5\tent{7}$ cm s$^{-1}$.}
  \label{fig:bf2-3wUy}
\end{figure*}

\subsubsection{Slow rotation - the case of \TF{}}
\label{sec:t5}

Here we discuss the special case of a star spinning at \nrep{11}, the
rotation rate of the source \TF{} discussed in \citet{\cael}. Recall
that in this case the Coriolis force is not capable of confining the
fluid. We ran simulations using the same fields as we used for the
case \nrep{450}, so that comparison will be easier, but the concept of
weak, intermediate and strong field should now be revisited: apart
from the case with \brep{1}{7}, all fields have frictional time scales
smaller than the rotational period and should be regarded as
\emph{strong}.

First, we ran another simulation, Run 8, with a seed magnetic field of
\brep{1}{10}, as in the case of Run 4. As expected, the acceleration
due to magnetic stresses dominates that due to the Coriolis force
therefore controlling the flame propagation and the simulation results
are almost indistinguishable from those at higher spin (Run
4). Therefore, also in this case, the flame propagation resembles that
confined only by the Coriolis force at higher frequency (see
\secref{sec:strong}) and thus we expect that a slow rotating source in
this regime of magnetic field will still have a phenomenology similar
to faster rotators \emph{without magnetic} field\footnote{The velocity
  reported in \tabref{tab:runs} is slightly lower than the one for Run
  4, but that is just a consequence of the fact that we stopped the
  simulation early. When overplotting the position versus time of the
  two simulations, they agreed almost perfectly.}. A simulation with
\brep{1}{9}, Run 7, also shows a behaviour very comparable to the case
with the same field strength and \nrep{450}, Run 3. That, again, can
be understood in terms of the frictional time scale (see
\tabref{tab:runs}). Furthermore, we noticed that the thermal wind
structure was absent at \nrep{450}, a fact that showed how the
Coriolis force was no longer playing a significant role for the
confinement and propagation.

\begin{figure*}
  \centering
  \doubfig{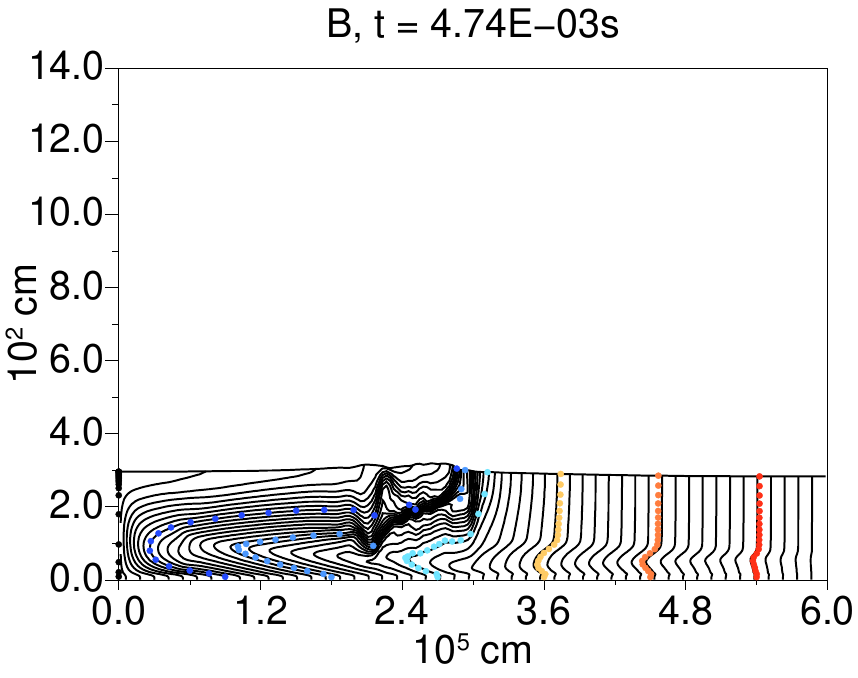}{}{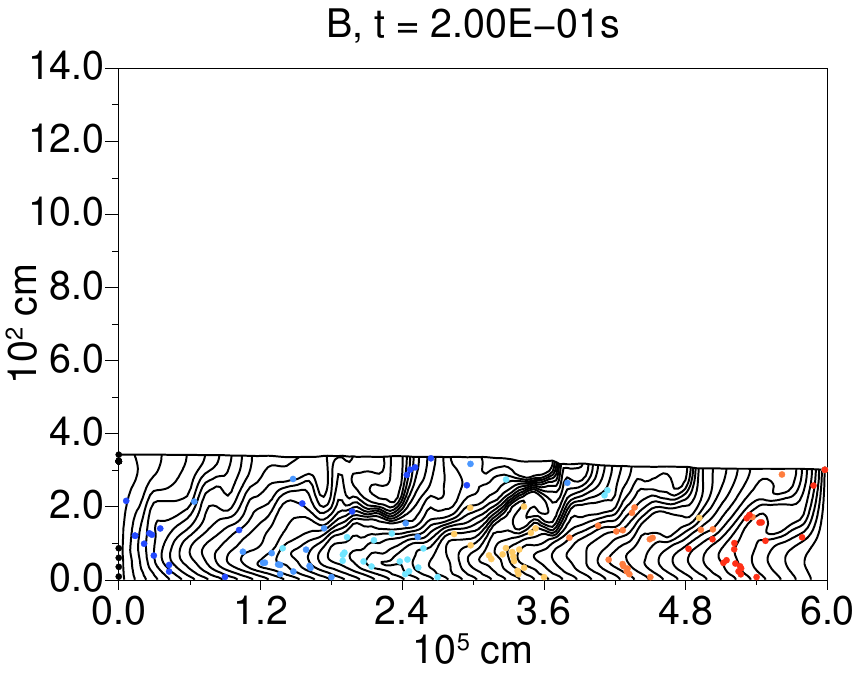}{}
  \doubfig{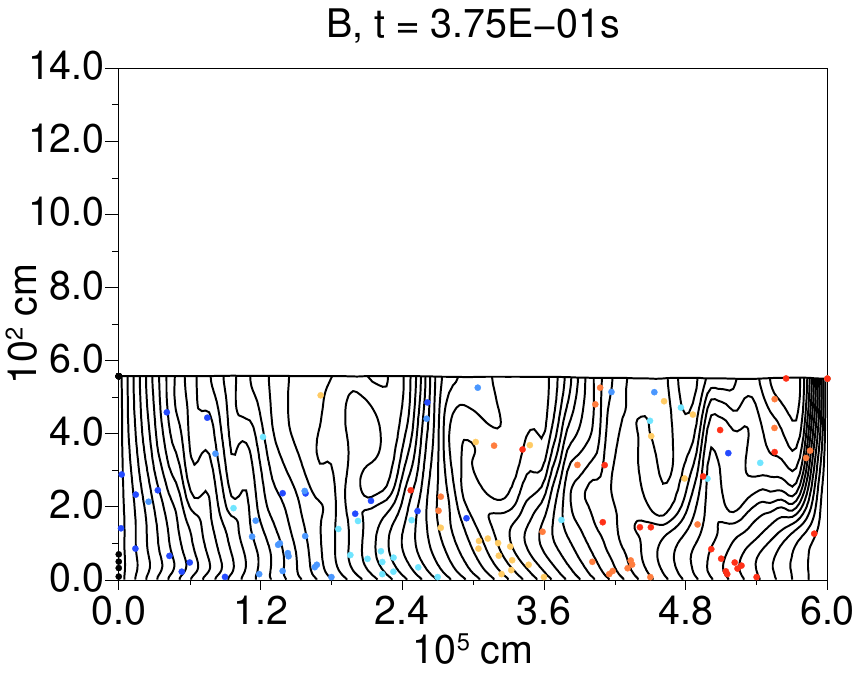}{}{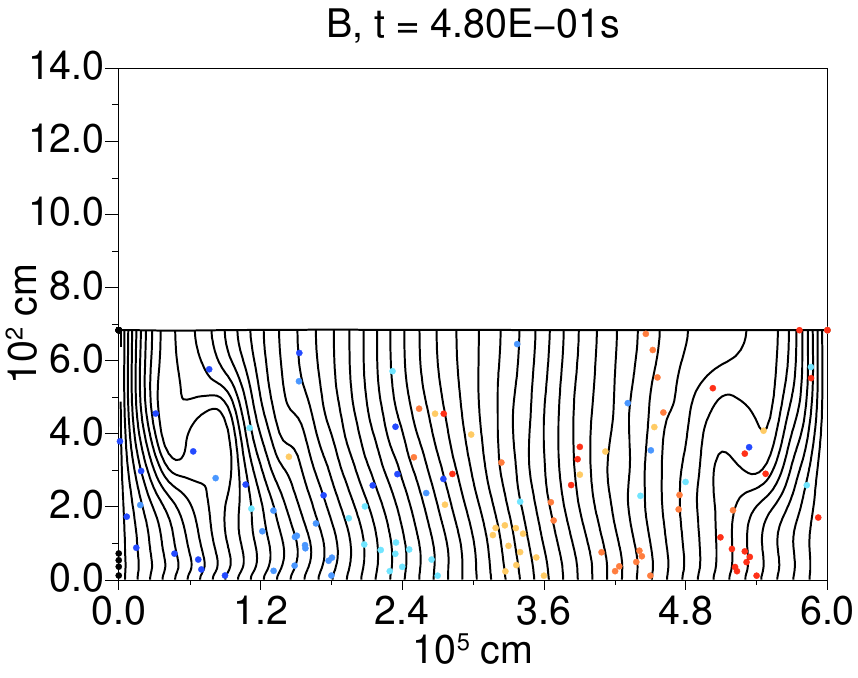}{}
  \caption{The evolution of Run 6, \brep{1}{8} \nrep{11}, at different
    stages. Shown are magnetic field lines with tracer particles
    superimposed. Particles with the same color had the same
    horizontal position at the beginning of the
    simulation. \textbf{Upper panels}: the fluid is confined by the
    magnetic field only (left). A very fast initial propagation takes
    the form of rolling eddies (right). \textbf{Lower panels}: later
    stages of the burst when ignition is almost horizontally
    homogeneous. Convection sets in (left) and eventually subsides
    (right).}
  \label{fig:bf9w}
\end{figure*}

When the field is \brep{1}{8}, Run 6, some differences between fast
and slow rotation become more noticeable. After an initial phase where
the fluid strongly bends the field lines, weak burning develops at the
left end of the simulation, which triggers rolls that propagate
towards the right similar to those of Run 2 (\brep{1}{8},
\nrep{450}). These rolling eddies compress the fluid lines and climb
over them at intervals. We see also fluid flowing down- and leftwards
along the fluid lines when the eddies hit a new region of compressed
lines.  This propagation is similar to the case with \brep{1}{8} and
\nrep{450}; however, at low rotation the length scale of these eddies
is obviously longer (\figref{fig:bf9w}, upper panels). This velocity of
propagation is very fast and at $t\sim2.6\tent{-1}$ s, the whole layer
is horizontally approximately homogeneously burning. It is difficult
to track a neatly defined flame front, since secondary ignition sites
ignite continuously due to the rolling eddies. Therefore, we do not
report any velocity in \tabref{tab:runs}, but a rough estimate gave
$\vf\sim3.5\tent{6}$ cm s$^{-1}$. Finally, when the burning is roughly
homogeneous we see a convective pattern with three or four cells above
the most intensely burning layer. These rolls disappear gradually
during the rise of the burst, leaving at the end only two rolls, one
per side, which eventually disappear as well, before the peak of the
burst (\figref{fig:bf9w}, lower panels).

Finally, in the case of \brep{1}{7}, Run 5, we see that the fluid
sloshes back and forth many times before ignition, which takes place
at $t\sim1.94$ s. This is to be expected, since neither the Coriolis
force nor the field (which is weaker) can significantly confine the
fluid within the simulation domain and the waves bounce back only
because of the reflecting boundaries. Without the magnetic field the
fluid behaved very similarly, starting ignition at around $t\sim1.77$
s. Convection develops above the horizontal layer that is burning,
similarly to the case with \brep{1}{8} and \nrep{11}.  Note that such
late ignitions are probably caused by the finiteness of our simulation
domain and the 2D dimensionality which prevents any dissipation in the
$\y$ direction, while a neutron star spherical surface is much wider
and in reality the flame would probably flume out. We conclude that
the results of our simulations for the case of \nrep{11} are in
agreement with the predictions of \citet{\cael}: a field of
$\Bo\gtrsim 10^9$ G is required in order to have flame confinement and
propagation.

\section{Summary and conclusions}
\label{sec:discussion}

\subsection{Summary of simulations at fast rotation (\nrep{450})}
\label{sec:summary}

In this paper we have presented 2D simulations of Type I bursts in the
presence of both rotation and an initially vertical magnetic field.
We have shown that in the case of a magnetic field of \brep{1}{10},
what we called the \emph{strong} magnetic field case, flame
propagation is independent of the rotation rates we used (\nrep{11},
\nrep{450}). The magnetic field prevents the development of any
thermal wind (the hurricane structure described in \citetalias{\slu}
and \citealt{\cath}) by inhibiting significant horizontal motions. The
flame structure resembles very closely the one of a fast rotator in
the sense that the fluid is not affected by turbulence, but
propagation is slower. On the other hand, we found that in the
presence of a \emph{weak} magnetic field (\brep{1}{7}), the fluid
confinement depends on the Coriolis force. In the case of fast
rotation the presence of the field affects the structure of the front,
making it more extended in the horizontal direction. This has the net
effect of speeding up the flame propagation compared to a case with
rotation only.

A striking feature we found in the case of fast rotation with
\brep{1}{7} is the development of a peak in the fluid height, due to
high temperature, at a location approximately coincidental with the
size of our initial heat reservoir in the temperature profile. A test
with different size of the perturbation confirmed that the position of
the peak is related to the initial conditions. From our simulations it
looks like the origin of higher temperature is due to extra heating
(mainly viscous) deposited by waves localized below the peak. Our
viscosity is numerical, but if a similar mechanism were at work in
nature, this would imply that there exist cases where a weak magnetic
field (when compared to the Coriolis force, in the sense defined in
\secref{sec:results-num}) would lead to the presence of an asymmetry
in the emission pattern during the rise of the burst. This, in turn,
could lead to the presence of burst oscillations. We also found that
the velocity of the flame seemed to be accelerating until it reached
the right boundary. Indeed, the flame propagation was better fitted by a
parabola than a straight line. It would be interesting to check
whether a constant velocity would ever be reached and on what length
scale. In particular, it would be interesting to follow the evolution
of this case on the spherical surface of a NS.

The case of intermediate strength magnetic field with \brep{1}{8} at
fast rotation showed a very remarkable series of results. First, we
saw a temperature and height peak propagating with the flame, which
would look like an expanding ring on the surface of the neutron star,
until the effects of curvature would become significant. The duration
and the possible effects on burst oscillations of this ring should be
tested in 3D simulations. Secondly, we saw a perturbation triggered by
the bending of the field lines at the very beginning of the
simulation. However, this did not prevent ignition and flame
propagation. Finally, this configuration showed that rolling eddies
form at the flame front, which mix the hot and cold fluid and speed up
the flame propagation by even an order of magnitude when compared to
the case without magnetic field. Also, similar eddies at the flame
front form in the case of a field \brep{1}{9}. In this case, however,
there is no trace of the thermal wind and the hurricane structure that
we still observe when \brep{1}{8} and \brep{1}{7}. The burst proceeds
with speed higher than in the case without magnetic field.

The presence of the eddies at the flame front is very interesting,
since they provide a means of heat conduction over a long length scale
helping in speeding up flame propagation as compared to when only
conduction is taking place \citep[see for example the discussion of
][about the different speeds in the case of conduction, convection and
  turbulence]{art-1982-fryx-woos-b}. These giant rolls could also be
responsible for bringing burning ashes to higher layers in the
atmosphere, a fact that could lead to the generation of absorption
lines \citep{art-2006-wein-bild-scha}. Their behaviour, in relation
with rotation rate and magnetic field strength, is very close to what
is expected for convective rolls in the standard \bena{} problem with
both rotation and magnetic field \citep{book-1961-chandra}. Indeed,
their size decreases with increasing magnetic field (in our cases,
going from \brep{1}{8} to \brep{1}{9}, for example) and increases with
decreasing rotation frequency (as clearly seen for the case with
\brep{1}{8} going from \nrep{450} to \nrep{11}). However, during the
bursts we do not have a standard \bena{} configuration, since the
fluid can be in motion due to the thermal wind structure (as in the
case of \brep{1}{8}), the flame front is continuously moving and so is
the source of heat that sets the temperature gradient. Furthermore,
the configuration at the front is affected by the baroclinic motions,
by the compression of the cold fluid under the weight of the expanding
hot fluid and the Parker instability. The rolls clearly have an
important role that could be amplified in 3D and further study is
required.

\subsection{Summary of simulations at slow rotation (\nrep{11}) and
  implications for the source \TF}
\label{sec:conctf}
We performed simulations of a star with \nrep{11} and magnetic fields
in the same range as for the fast rotator case: $10^7 \le \Bo \le
10^{10}$ . Simulations with \brep{1}{10} and \brep{1}{9} behaved
practically identically to the cases with \nrep{450}, since the
magnetic field was providing most of the confining force also in the
case of fast rotation. The case with \brep{1}{8} showed similar eddies
to the analogue case at faster rotation, even though the length scale
of the eddies was longer. The initial phase of flame propagation was
even faster and eventually the explosion of the burst was mostly
homogeneously horizontal. In the case with \brep{1}{7} neither the
magnetic tension nor the Coriolis force was strong enough to provide
any confinement and this led to the dispersion of the initial heat
reservoir over the full horizontal extent of the simulations. We still
saw a very delayed ignition, but we think that this is due to the
smallness of our simulation domain compared to the size of the neutron
star surface, a fact which prevents a more realistic dissipation.

We note that our simulations for the case of \nrep{11} confirm the
conclusions of \citet{\cael} about the case of \TF. This source was
discovered in 2010 in the globular cluster Terzan 5
\citep{atel-2010-bordas-etal}. It was found that \TF{} displayed both
accretion powered oscillations and burst oscillations at \nrep{11}
\citep{atel-2010-stroh-mark,atel-2010-alta-etal,art-2011-papit-etal}.
The accretion powered and the burst oscillations had highly accurately
matching phases and frequencies \citep{\cael}. This matching and the
slow rotation rate led the authors to exclude the possibility that the
Coriolis force could confine the flame and they also concluded that
neither a Coriolis force confined hot-spot nor global modes could
explain the burst oscillations. Based on approximated analytic
calculations the authors estimated that the magnetic field of \TF{}
could provide a dynamically effective confining force for the flame if
$B\gtrsim 10^9$ G.

Therefore, a field weaker than \brep{1}{8} would not confine the fluid
and both ignition and burst oscillations should not be expected.  At
\brep{1}{8}, confinement is still weak and even if we do see the
explosion leading to a burst, we do not expect to see any burst
oscillation at this regime, since the flame propagates very fast and
the whole domain ignites almost coincidentally. On the other hand, a
field of \brep{1}{9} or higher shows a confined propagating flame, and
we conclude that this is the minimum field intensity needed to have
bursts that could display burst oscillations.

\subsection{Discussion of numerical and nuclear burning effects}
\label{sec:conv}

Before tackling the effects of the magnetic field, we will briefly
discuss possible contributions that may affect the flame velocity we
measure.

One possible factor is resolution-related numerical effects. We could
not perform full convergence tests for the simulations with magnetic
field since that would have consumed many months, an amount of
computational resources that we do not have at the moment. However,
the convergence tests in \citet{\cath} showed that our code at similar
resolutions to those used in this paper is converging, even if
somewhat slowly.  Furthermore, short convergence tests for the initial
stages\footnote{The extension of the tests much further that the
  initial stages is prevented by the limitations on computing time.}
of runs $0 - 4$ showed the following trends.

Run 0, being purely hydrodynamical, behaves exactly as the tests
performed in \citet{\cath} with the velocity of the flame decreasing
slowly with resolution; run 4 behaves similarly, but the velocity
\emph{at this stages} seems already almost converged (velocity changed
by only $\sim 5$ \%). In the case of run 1 we note that at resolutions
much lower than those in the text, many of the waves associated with
the flame would not be resolved. The velocity measured at this very
initial stages is \emph{increasing} with resolution, by $\sim 30$ \%,
but the velocity we report in the text is evaluated at later stages
that we did not reach in this short convergence tests. Runs 2 and 3,
on the other hand, should be more sensitive to resolution. This is
linked to the fact that (a) the maximum flame position is difficult to
track for both simulations and (b) the convective rolls are captured
differently at different resolutions. This situation is complicated by
the fact that at different resolution our numerical diffusivities are
different \citep{\cath}.  As expected, run 2 is more affected by
resolution from the beginning, but the qualitative behaviour is the
same as in our reference run, with the development of the initial
perturbation described in \secref{sec:intermediate}. The velocity in
run 3 seems almost converged in the early stages (difference in
velocity is $\sim 0.01$ \%), but we cannot exclude that it might
change at later stages due to the onset of rolls (see
\secref{sec:intermediate}).

Other important factors are the physical burning and the
opacity. \citet{\timm} performed a detailed analysis of 1D horizontal
laminar propagation of the flames, studying its dependence on the
reaction network, thermodynamic conditions and
conductivity. \citeauthor{\timm}' simulations were in good agreement
with order of magnitude calculations predicting a velocity of the
flame $\propto \sqrt{Q/\kappa}$, with $Q$ the burning rate and
$\kappa$ the opacity. A direct comparison to his results is
impossible, mainly because the flame in our simulations, being
vertically resolved, crosses different levels of pressure, while
\citeauthor{\timm}' simulations were performed at constant pressure.
Furthermore, we include cooling and fluid dynamics, which are absent
in that analysis. In \citet{\cath} we proved that flame propagation in
type I bursts can be expressed as a vertical flame velocity $\vperp$
times a geometrical boosting factor set by the slope of the hot-cold
fluid interface: $\vf \sim \vperp \ro / H$. As we showed in the
present paper, in many cases the net effect of the magnetic field is
to interfere with the Coriolis force, therefore changing the slope of
the interface and the boosting factor. Since this is the focus of the
paper, we discuss it in the next section. However, a different
conduction or a different burning regime as used by \citeauthor{\timm}
will change the vertical flame speed $\vperp$. Our conduction is
constant across the whole domain. While this is an approximation, it
is less important than the reaction rate, since the value we use is
calibrated to be the expected opacity at the conditions near the most
burning regions. On the other hand, our reaction rate is limited to
helium burning into carbon. A more extended reaction network with
quick energetic reactions will boost the propagation speed by
increasing the energy production rate at the \emph{vertical} flame
front. That is because of the flame speed dependence $\vperp \propto
\sqrt{Q}$ \citep[][]{\timm}. Slow reactions, like for example those
limited by $\beta$-decays in hydrogen rich material, will probably
have a smaller impact as long as there are faster ones occurring,
since the slow ones will take place in the wake of the vertical flame
front. Direct numerical simulations are needed to better quantify
these effects in the complex dynamical configurations of the type I
bursts, in order to quantify the effects of composition and nuclear
burning on the propagation speed and surface temperature distribution
during all the stages of the bursts. This will also have a direct
impact on the prediction and measurement of burst oscillations.

\begin{figure*}
  \centering
  \doubfig{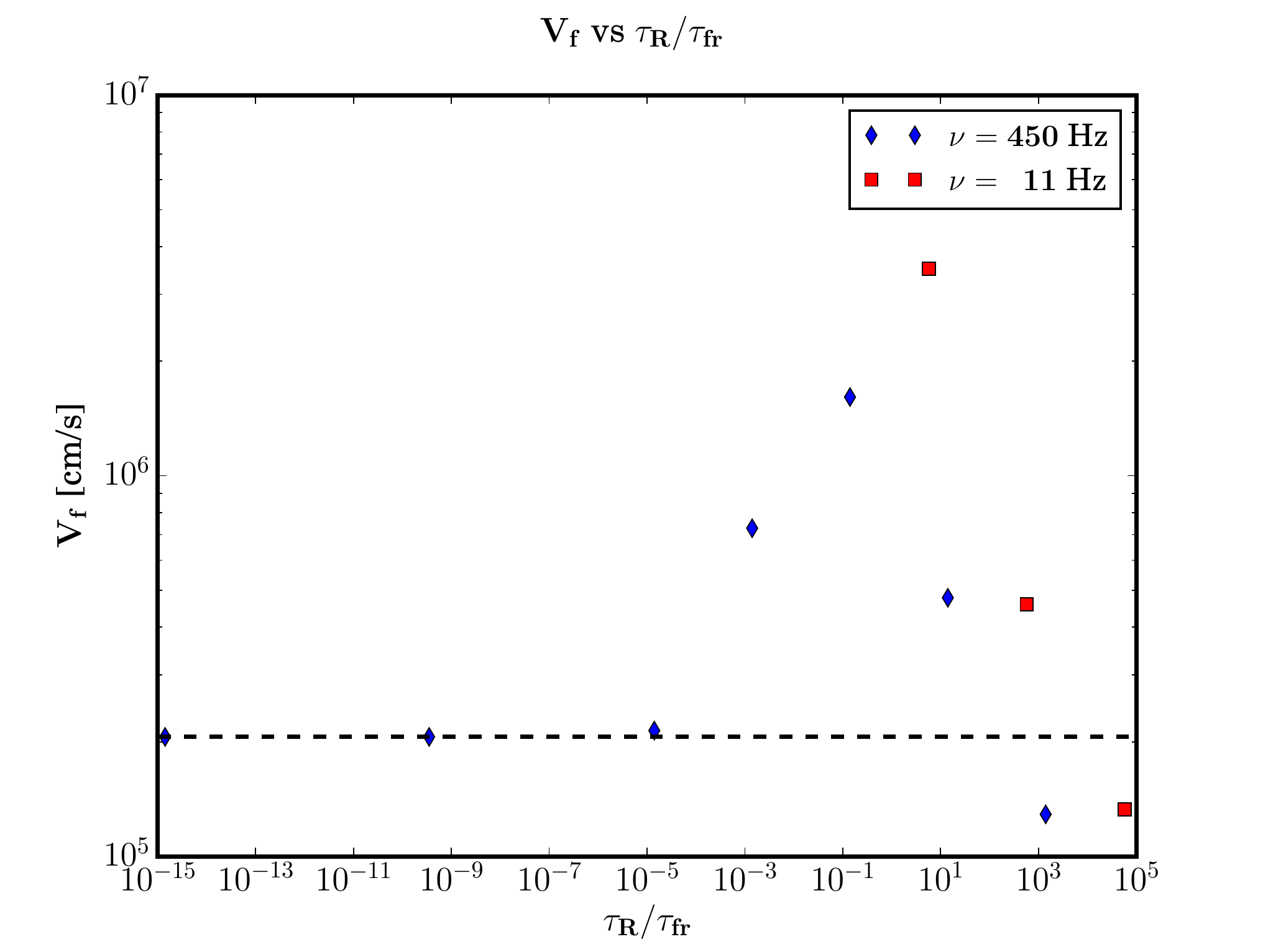}{}{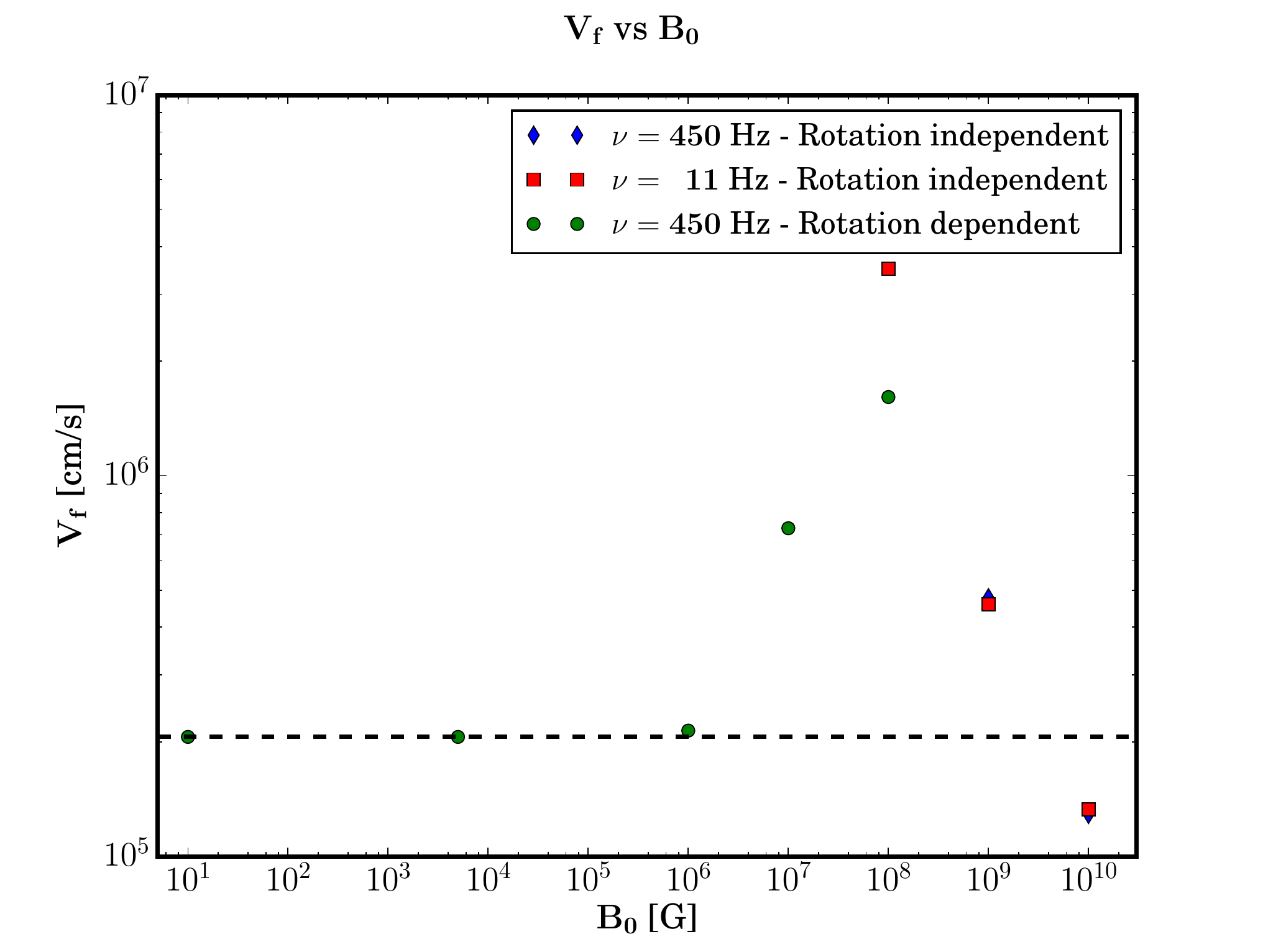}{}
    \caption{\textbf{Left}: velocity of the flame versus
      $\impt$. The blue diamonds represent the results for the fast
      rotator \nrep{450}, while the red squares are the results for
      the slow rotator case (\nrep{11}). The horizontal dashed line
      indicates the velocity of the flame when magnetic field is
      absent. \textbf{Right}: velocity of the flame versus $\Bo$ (note
      that $\Bo \propto 1/\tb \propto 1/\sqrt{\tf}$). The blue
      diamonds represent the results for the fast rotator \nrep{450}
      and the red squares are the results for the slow rotator
      (\nrep{11}) when $\tr\gg\tf$ and the flame velocity is
      independent of rotation. The green circles are the results for
      the fast rotator when $\tr<\tf$ and the velocity is affected by
      the Coriolis force. In these plots we include results both from
      \tabref{tab:runs} and footnote \thefoota.}
  \label{fig:VB}
\end{figure*}

\subsection{The effect of the magnetic coupling on flame propagation speed}
\label{sec:frictional}

The most remarkable result of our simulations is about the effect that
the magnetic field has on the propagation speed of the flame. We
summarize the results regarding the flame speed $\vf$ in
\figref{fig:VB}. Note that we include the values both from
\tabref{tab:runs} and footnote \thefoota{} (compare also to figure 2
of \citetalias{\slu}).

In \secref{sec:results-calc} we argued that there exists a field
strength at which the propagation speed is maximal (\equlab{.}
\ref{BZcr}, $\Bo\sim2.8\tent{8}$ G for \nrep{450}). We also discussed
how for decreasing field strength less effective coupling will lead to
lower values of the flame speed, to the point that when $\tr\ll\tf$
the velocity should be independent of the magnetic field
(\equlab. \ref{vlow}). This is the trend we see in the left panel of
\figref{fig:VB} looking at the results for \nrep{450} (blue
diamonds). If $\Bo\gg\bar{\Bo}$, $\tr\gg\tf$, the flame speed is
expected to decrease continuously with the field strength. Again, this
can be verified in \figref{fig:VB}.

We also noted that when $\tr\gg\tf$ the velocity of propagation should
depend only on the field strength via the frictional time scale $\tf$
(\equlab. \ref{vhigh}). This is more clearly seen in the right panel
of \figref{fig:VB}. There, the blue diamonds (\nrep{450}) and the red
squares (\nrep{11}) represents the results obtained in this regime. In
particular, it can be seen that the velocities for \brep{1}{9} and
\brep{1}{10} overlap almost perfectly. In contrast, for $\Bo\leq10^8$
G the results at \nrep{450} deviate from those at \nrep{11} because of
the increasing importance of the Coriolis force on the formers (green
circles), as expected.

We conclude that our simulations confirm our analysis on the
importance of the magnetic field in regulating the flame speed by
providing an effective frictional coupling among the fluid layers and
we note that for realistic field strengths we obtain speed values that
are already in very good agreement with the velocities inferred from
observations.\\

{\noindent\it Acknowledgements.} We thank Frank Timmes for making his
astrophysical routines publicly available. Y.C. and A.W. acknowledge
support from an NWO Vrije Competitie grant ref. no. 614.001.201
(PI: Watts). This work was also sponsored by NWO Exacte Wetenschappen
(Netherlands Organization for Scientific Research, Physical Sciences)
for the use of supercomputer facilities, with financial support grant
no. SH-337-15 (PI: Cavecchi). The paper benefitted from NASA's
Astrophysics Data System.

\bibliographystyle{mnras}
\bibliography{ms}
\label{lastpage}

\end{document}